\documentclass[aps,prb,twocolumn,eqsecnum,amsmath,amssymb,floatfix,superscriptaddress]{revtex4}
\usepackage{tabularx}
\usepackage{subfigure}
\usepackage{subfigure}
\usepackage{graphicx}
\usepackage{color}

\def\ie{{\it i.e.,\ }}

\def\slD{\raise.15ex\hbox{$/$}\kern-.57em\hbox{$D$}}
\def\dsl{\raise.15ex\hbox{$/$}\kern-.57em\hbox{$\Delta$}}
\def\slp{{\raise.15ex\hbox{$/$}\kern-.57em\hbox{$\partial$}}}
\def\nsl{\raise.15ex\hbox{$/$}\kern-.57em\hbox{$\nabla$}}
\def\sla{\raise.15ex\hbox{$/$}\kern-.57em\hbox{$\rightarrow$}}
\def\slla{\raise.15ex\hbox{$/$}\kern-.57em\hbox{$\lambda$}}
\def\slb{\raise.15ex\hbox{$/$}\kern-.57em\hbox{$b$}}
\def\lnp{\raise.15ex\hbox{$/$}\kern-.57em\hbox{$p$}}
\def\lnk{\raise.15ex\hbox{$/$}\kern-.57em\hbox{$k$}}
\def\lnK{\raise.15ex\hbox{$/$}\kern-.57em\hbox{$K$}}
\def\lnq{\raise.15ex\hbox{$/$}\kern-.57em\hbox{$q$}}
\def\lnA{\raise.15ex\hbox{$/$}\kern-.57em\hbox{$A$}}
\def\lna{\raise.15ex\hbox{$/$}\kern-.57em\hbox{$a$}}
\def\lnB{\raise.15ex\hbox{$/$}\kern-.57em\hbox{$B$}}
\def\slA{\raise.15ex\hbox{$/$}\kern-.57em\hbox{$A$}}
\def\slash#1{\raise.15ex\hbox{$/$}\kern-.57em\hbox{$#1$}}

\def\ket#1{\vert#1\rangle}      

\begin{document}

\title{Quantum criticality, lines of fixed points, and phase separation
in doped two-dimensional quantum dimer models}

\author{Stefanos Papanikolaou}
\affiliation{Department of Physics, University of Illinois at
Urbana-Champaign, 1110
W. Green St., Urbana, Illinois 61801-3080}
\author{Erik Luijten}
\affiliation{Department of Materials Science and Engineering, University of
Illinois at Urbana-Champaign, 1304 W. Green St., Urbana, Illinois 61801-2920}
\author{Eduardo Fradkin}
\affiliation{Department of Physics, University of Illinois at
Urbana-Champaign, 1110
W. Green St., Urbana, Illinois  61801-3080}

\date{\today}

\begin{abstract}
  We study phase diagrams of a class of doped quantum dimer models on the square
  lattice with ground-state wave functions whose amplitudes have the form of the
  Gibbs weights of a classical doped dimer model. In this dimer model,
  parallel neighboring dimers have attractive
  interactions, whereas neighboring holes either do not interact or have
  a repulsive interaction.  We investigate the behavior
  of this system via analytic methods and by Monte Carlo
  simulations.  At zero doping, we confirm the existence of a
  Kosterlitz-Thouless transition from a quantum critical phase to a
  columnar phase. At low hole densities we find a dimer-hole liquid phase and a
  columnar phase, separated by a phase boundary which is a line of critical points with varying exponents.
  We demonstrate that this line ends at a multicritical point where the
  transition becomes first order and the system phase separates. The
  first-order transition coexistence curve is shown to become unstable
  with respect to more complex inhomogeneous phases in the presence of
  direct hole-hole interactions. We also use a variational approach to determine the spectrum of low-lying density fluctuations in the dimer-hole fluid phase.

\end{abstract}
\maketitle

\section{Introduction}
\label{sec:intro}

The behavior of doped Mott insulators is a long-standing open and
challenging problem in condensed-matter physics. Mott insulators are the parent states of all strongly correlated electronic systems and as such play a crucial role
in our understanding of high-$T_c$ superconductors (HTSC) and many other
systems.  Their strongly correlated nature implies that their behavior cannot be
understood in terms of weakly coupled models. Except for the very special case of one spatial
dimension, the physics of doped Mott insulators is currently only understood
at a qualitative level. The solution of this challenging
problem remains one of the most important directions of research in
condensed-matter physics.

Quantum dimer models (QDM)\cite{rokhsar88} provide a simplified, and
rather crude, description of the physics of a Mott insulator. They provide a correct description of the physics of Mott
insulators in regimes in which the spin excitations have a large spin
gap.  QDMs were proposed originally within the context
of the resonating-valence bond (RVB) mechanism of
HTSC.\cite{anderson87,kivelson87} These systems are of great interest as they can yield hints on the
behavior of more realistic models of quantum
frustration. 

The main idea behind the formulation
of QDMs is that, if the spin gap is large, the spin degrees of
freedom become confined in tightly bound singlet states which, in the
extreme limit of a very large spin gap, extend only over distance scales of
the order of nearest-neighbor sites of the lattice. Thus, in this
extreme regime, the Hilbert space can be approximately identified (up to
some important caveats\cite{rokhsar88}) with the coverings of the
lattice by valence bonds or dimers.

Surprisingly, even at the
level of the oversimplified picture offered by QDMs, the physics of doped Mott insulators remains poorly understood.
  In this paper we explore the phase diagrams, and the properties of their phases, of QDMs generalized to include interactions
between dimers (or valence bonds), and between dimers and doped
(charged) holes. Basic aspects of the physics of these models are reviewed in
Ref.~\onlinecite{fradkin91} and references therein.

Undoped QDMs have been studied more extensively and by now they are relatively well
understood.\cite{read-sachdev91,sachdev-read91,moessner01a,moessner02a,fradkin04,vishwanath04,ardonne04}
On bipartite lattices, their ground states show either long-range
crystalline valence bond order of different sorts or are quantum
critical,\cite{rokhsar88,fradkin91,moessner02a} while on non-bipartite
lattices their ground states are typically disordered and are topological fluids.\cite{moessner01a} 

The
more physically relevant \emph{doped} QDMs, with a finite density of
charge carriers (holes), are much less understood, although some
properties are
known.\cite{fradkin90a,balents05b,alet05,syljuasen05,alet06,Castelnovo06,Poilblanc06}
In QDMs a spin-$\frac{1}{2}$ hole fractionalizes into  a bosonic {\em
holon}, an excitation that carries charge but no spin, and a {\em
spinon}, a fermionic excitation that carries spin but no
charge.\cite{kivelson87,rokhsar88,fradkin90a,fradkin91} Holons can be
regarded as sites that do not belong to any dimer, whereas spinon pairs are broken dimers. This form of electron fractionalization is observable in the spectrum of these systems only in the topological disordered (spin-liquid) phases of the undoped QDM. Otherwise, as in the case of the valence bond crystalline states which exhibit long range dimer order, spinons and holons are {\em confined} and do not exist as independent excitations.\cite{fradkin91}

In this paper, we consider several interacting QDMs on a square
lattice at finite hole doping, and discuss their possible phases and
phase transitions as a function of hole density and strength of the
interactions. At any finite amount of doping the system will have a finite density of holes, which are hard core charged bosons in this description. To simplify the problem, in this work we do not consider the physical effects of the charge-neutral fermionic spinons which in principle should also be present.  Thus, at this level of approximation, all spin-carrying excitations are effectively projected out. The remaining degrees of freedom are thus dimers (``spin-singlet valence bonds'') and charged hard-core bosonic holes. Already this simplified picture of a strongly correlated system is very non-trivial.

For a certain
relation between its coupling constants, known as the Rokhsar-Kivelson (RK)
condition,\cite{rokhsar88} QDM Hamiltonians, both with and without holes,  can be written as a sum
of projection operators. These RK Hamiltonians  are manifestly positive definite operators. For this choice of couplings, the ground-state
wave function is a  zero-energy state which is known exactly. This {\em RK wave function} is a local function of the degrees
of freedom of the QDM, the local dimer and hole densities. The quantum-mechanical
amplitudes of the RK wave functions turn out to have the same form as the Gibbs weights
of a two-dimensional (2D) \emph{classical} dimer problem with a finite density of holes.
For the generalized doped QDMs that we consider, the norm of the exact
ground-state wave function is equal to the partition function of a
system of interacting classical dimers at finite hole density. This
mapping to a 2D classical statistical mechanical system, for which there
is a wealth of available results and methods, makes this class of
problems solvable.\cite{rokhsar88,moessner01a,ardonne04,Castelnovo06} 

In this work we will investigate the behavior of doped QDMs which satisfy the RK conditions by studying the correlations encoded in their ground state wave functions.
The phase diagrams of these systems turn out to be quite rich.  As we shall
see, these simple models can describe many aspects of the physics of
interest in strongly correlated systems, including a dimer-hole liquid
phase (a Bose-Einstein condensate of holes), 
valence-bond crystalline states, phase separation, and more general
inhomogeneous phases.  The undoped version of this system was studied in detail in
Ref.~\onlinecite{alet05}, where a quantum phase transition was found
that was argued to belong to the Kosterlitz-Thouless (KT) universality
class, from a critical phase to a columnar state with long-range order.
In this paper we confirm that this is indeed the case. At finite hole
density, hitherto available results are limited to the form of the
associated RK QDM Hamiltonian\cite{Castelnovo06,Poilblanc06} and
numerical results for small systems.

In this work we employ analytic
methods\cite{kadanoff77,ginsparg88,boyanovsky89a,lecheminant02} combined
with advanced classical Monte Carlo (MC) simulations\cite{krauth03,liu04} to
probe the correlations in the {\em doped} RK wave functions, investigate the phase
diagram and its phase transitions. The methods used here can be readily
generalized to the case of non-bipartite lattices, for which a number of
important results have been
published.\cite{moessner01a,fendley02,trousselet07} In Section
\ref{sec:hamiltonians}, 
we describe the construction of two generalized quantum dimer RK
Hamiltonians that we used in our study. A similar but independent
construction has been presented by Castelnovo and coworkers\cite{Castelnovo06}
and by Poilblanc and coworkers\cite{Poilblanc06}. The RK wave functions of these generalizations of the
quantum dimer model have either a
fixed number of holes or a variable number of holes and a fixed hole
fugacity.  The ground-state wave functions of both models at their
associated RK points  correspond to a canonical dimer-monomer system in
the canonical and grand-canonical ensembles respectively. Near the end
of the paper, in Section \ref{sec:hole-int}, we 
introduce a third Hamiltonian, with an associated RK wave function, to
study the effects of hole interactions which compete with phase
separation at the first-order transitions that we find for both models.

In Sections \ref{sec:mean-field}
and~\ref{sec:exact-critical-behavior} we  study the correlations and the
phase diagram for the ground states encoded in these wave functions by
means of an analysis  
of the equivalent classical statistical system of dimers and holes for the RK Hamiltonians of  Section
\ref{sec:hamiltonians}. In Section
\ref{sec:mean-field} we summarize the results of a mean-field theory for
both non-interacting and interacting classical dimer models at finite
doping. The details of the mean-field theory are presented in Appendix
\ref{app:mean-field-details}, where we compute the hole-hole correlation
function and derive a qualitative phase diagram as a function of hole
density and dimer interaction parameters. The main result of this simple
mean-field theory is that the phase diagram at finite hole density
contains two phases, a dimer-hole fluid and a columnar dimer
solid. The columnar-liquid transition is continuous at weak coupling and
turns first order at a tricritical point. Naturally, the critical and
tricritical behavior are not correctly described by the mean-field
theory, although the general topology of the phase diagram is correct
and, remarkably, even the location of the tricritical point is
consistent with what we find in the MC simulations of Section~\ref{sec:MC}.

In Section~\ref{sec:exact-critical-behavior} we present a detailed analytic theory
of the critical behavior of interacting classical dimers. Sections \ref{sec:zero-density}
and~\ref{sec:finite-density} focus on the field-theoretic Coulomb-gas
approach for this model at zero and finite
doping, respectively. We show that, up to a critical value of a parameter, the undoped RK wave
function describes a critical system with continuously varying critical
exponents, with a phase transition (belonging to the 2D KT universality class) to
a state with long-range columnar order. At finite hole doping we find a
hole-dimer liquid phase (with short-range correlations) and a stable
phase with long-range columnar order. At low hole densities the phase
boundary is a line of fixed points with varying exponents ending at a
KT-type multicritical point where the transition becomes first order.
We present a field-theoretical treatment of this tricritical point and 
a theory of the evolution of the behavior of the columnar and
orientational order parameters and of their susceptibilities along the
phase boundary.  Past the tricritical point the system is found to
exhibit a strong tendency to phase separation, which we verify in our
numerical simulations (Section~\ref{sec:MC}). In Section~\ref{sec:hole-int} we 
consider the effects of direct hole-hole interactions near the
first-order phase boundary, and discuss one of the many
inhomogeneous phases which arise in this regime instead of phase
separation.

In Section \ref{sec:MC} we confirm our analytic predictions via
extensive classical MC simulations of the generalized RK wave
functions.  For the study of the line of critical points at low doping
we employ the canonical generalized geometric cluster algorithm (GGCA),
whereas the first-order transition is studied via grand-canonical
Metropolis-type simulations.  The GGCA algorithm enables us to study
relatively large systems, up to $400 \times 400$, for a range of
dopings, $0 \leq x \leq 0.06$, and to investigate the finite-size
scaling behavior. The accessible range of system sizes should be
compared to what can be reached for full quantum models, away from the
RK condition, where available methods, such as exact diagonalization
and Green function Monte Carlo, allow the study of only very small
systems with few holes.\cite{syljuasen05,Poilblanc06} 

In the undoped
case we confirm the existence of a Kosterlitz-Thouless transition from a
line of critical points to an ordered columnar state, as found in the
work of Alet and coworkers.\cite{alet05,alet06} We study the scaling
behavior of the columnar and orientational order parameters and of their
susceptibilities.  We also use a mapping of the orientational order
parameter of the interacting classical dimer model to the staggered
polarization operator obtained by Baxter for the six-vertex model\cite{baxter73} to fit
our MC data and find an accurate estimate of the KT transition
coupling in the undoped case.

At low doping, we study the transition from the dimer-hole fluid
phase to the columnar state. We confirm that the scaling dimension of
the columnar order parameter operator is equal to $1/8$, as predicted by
our analytic results of Section \ref{sec:finite-density}. We also
present a typical set of data that demonstrates how the scaling dimension of
the orientational order parameter varies, again in agreement with the
analytic results of Section \ref{sec:finite-density}, and use these
results to locate numerically the phase boundary. We then turn to the
behavior at larger doping and stronger couplings where the transition
becomes first order. We study this regime using grand-canonical
Metropolis Monte Carlo simulations. We confirm the first-order nature of
the phase transition by means of a careful analysis of the finite-size scaling
behavior of the order parameters across the phase boundary and of their
susceptibilities. We use these results to locate the phase boundary in
the first-order regime as well. In Section \ref{sec:hole-int},
we use MC simulations to study the effects of a direct hole-hole
repulsion which suppresses the effects of phase separation, leading instead to a complex phase diagram of
inhomogeneous phases, of which we only study its most commensurate case.

In Section~\ref{sec:SMA}, we study the elementary quantum excitations of
the doped QDMs satisfying the RK condition using the single-mode
approximation. We only present the main results and have relegated the
details to Appendix~\ref{app:fks}. We find that in the dimer-hole liquid
phase, hole and dimer density fluctuations have quadratic dispersions
$E(k)\sim k^2$. Thus this phase should be characterized as a
Bose-Einstein condensate of bosonic charged particles (holes), but not really
a superfluid, for reasons similar to those of Rokhsar and
Kivelson.\cite{rokhsar88}
We summarize our overall conclusions in Section
\ref{sec:conclusions}.

While this manuscript was being completed (and refereed) a number of
independent studies of aspects of this problem have been
published.\cite{Castelnovo06, Poilblanc06,alet06} Our results agree with
those in these references wherever they overlap, as noted throughout
this paper.
 
\section{Quantum Hamiltonians for Interacting Dimers at Finite Hole Density}
\label{sec:hamiltonians}
 
The Hamiltonian of the quantum dimer model (QDM) can be written in the
Rokhsar-Kivelson (RK) form\cite{rokhsar88} as the sum of a set of
mutually non-commuting projection operators $Q_p$,
\begin{equation}
 H=\sum_{\{p\}} Q_p \;,
\end{equation}
where $\{p\}$ denotes the set of all plaquettes of the square lattice.
Each projection operator $Q_p$ acts on the states of the dimers and
holes of a plaquette $p$ (or set of plaquettes surrounding $p$). In the
simplest case\cite{rokhsar88} each $Q_p$ acts only on the states labeled
by the dimer occupation numbers of the links of the plaquette $p$. In
this case, the ground state is described by the short-range RVB wave
function\cite{kivelson87}
\begin{equation}
 \ket{\textrm{RVB}}=\sum_{ \{ \mathcal{C} \} } \ket{\mathcal{C}} \;,
\end{equation}
where $\{ \mathcal{C} \} $ is the set of  (fully packed) dimer coverings
of the 2D square lattice, and $\{ \ket{\mathcal{C}} \}$ is a complete
set of orthonormal states. If one regards the dimers as spin singlet
states (with the spins residing on the lattice sites) each configuration
represents a set of spin singlets or
valence-bonds.\cite{anderson87,kivelson87} The dimer representation
ignores the over-completeness of the valence bond singlet states.\cite{rokhsar88} This problem can be made parametrically
small using a number of schemes, including large $N$ approximations\cite{read-sachdev89} and decorated generalizations of the spin-$1/2$ Hamiltonians.\cite{Raman05}
 
It is possible and straightforward to generalize the QDM construction so as
to include other types of interactions and coverings.  In
Ref.~\onlinecite{ardonne04} it was shown how to extend this structure to
smoothly interpolate between the square and the triangular lattices.  It
was also shown there that the same ideas can be used to construct a
quantum generalization of the two-dimensional classical Baxter (or
eight-vertex) model. In all of these cases, the RK form of this
generalized
quantum dimer model has an exact ground-state wave function whose
amplitudes are equal to the statistical (Gibbs) weights of an associated
two-dimensional classical statistical mechanical system on the same
lattice. Thus, if the classical problem happens to be a classical
critical system, the associated wave function now describes a 2D problem
at a quantum critical point. In Ref.~\onlinecite{ardonne04} such
quantum critical points were dubbed ``conformal quantum critical
points'' since the long-distance structure of their {\em ground-state
wave functions} exhibits 2D
conformal invariance. Here we are interested in a different
generalization of the QDM in which we consider dimer coverings
(although not necessarily fully packed) of the square lattice. We will
also consider 2D Hamiltonians whose wave functions correspond to
classical interacting 2D dimer problems with local weights. Similar but
independent constructions have also been proposed.\cite{castelnovo05,Castelnovo06,Poilblanc06}

Trying to be as physical and local as possible, we keep the
quantum-resonance terms as simple as before (single plaquette moves), but the
potential terms (which again have a central plaquette\cite{kivelson87}) now
have fine-tuned couplings that depend on the nearby plaquettes. Explicitly, we
have (cf.\ Fig.~\ref{dimerconfig}):

\begin{figure}[t]
\includegraphics[width=0.48\textwidth]{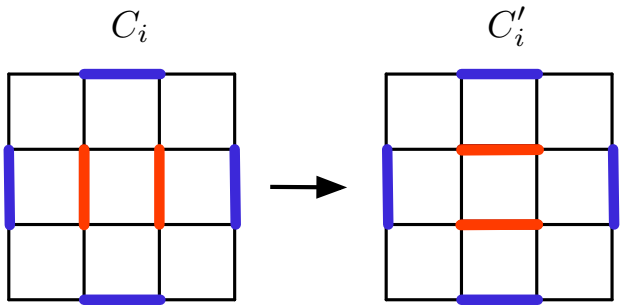}
\caption{(color online) Illustration to clarify the construction of the
Hamiltonian~\eqref{Ham}. The central dimer pair is present in all
configurations $C_i$ and $C_i'$, whereas the four surrounding dimers may or
may not be present. The index~$i$ enumerates the $2^4$ possible arrangements
of the surrounding dimers. The configuration $C'$ differs from the
corresponding $C$ by a single flip resonance of the central dimer pair.}
\label{dimerconfig}
\end{figure}

\begin{eqnarray}
  H_{d} &=& t\sum_i \Bigg[- \Big|C_i\Big>\Big<C_i'\Big| - \Big|C_i'\Big>\Big<C_i\Big| \nonumber\\
  &+&
  w^{R_{C_i'}-R_{C_i}}\Big|C_i\Big>\Big<C_i\Big|+w^{R_{C_i}-R_{C_i'}}\Big|C_i'\Big>\Big<C_i'\Big|\Bigg] \;, \nonumber \\
  &&
\label{Ham}
\end{eqnarray}
where $R_{C_i}$ and $R_{C_i'}$ denote the number of pairs of present
dimers formed in configurations $C_i$ and $C_i'$ respectively.

The Hamiltonian of Eq.~\eqref{Ham} is designed in such a way that it
annihilates any superposition of dimer-configuration states which have
amplitudes that are of the form $w^{N_p}$, where $w$ is the
parameter appearing in the Hamiltonian and $N_p$ is the number
of pairs of neighboring dimers in the configuration.\cite{ardonne04} In
this sense, the Hamiltonian is a sum of projection operators and
consequently there is a unique ground state for each topological sector
which must be composed of the superposition of these specially weighted
configurations.  
The ground-state wave function, $\ket{G}$, the state annihilated by all the projection operators, 
for this system is:
 \begin{equation}
\ket{ G}=\frac{1}{\sqrt{Z(w^2)}}\sum_{\{\mathcal{C}\}} w^{\displaystyle{N_p[\mathcal{C}]}} \ket{\mathcal{C}} \;,
\label{Gzw}
\end{equation}
where $Z(w^2)$, the normalization of this state,
\begin{eqnarray}
 Z(w^2) =\sum_{\{\mathcal{C}\}} w^{\displaystyle{2N_p[\mathcal{C}] } }
 \label{z1}
\end{eqnarray}
has the form the partition function of classically interacting dimers with a
coupling $u=-2\ln w$ between parallel neighboring dimers. In the
following, we will assume an attractive coupling,  $u<0$ or $w>1$. The case
$u>0$ was studied for the fully-packed case in
Ref.~\onlinecite{Castelnovo06}.

There are two different ways in which we can add doping to our system, while
still being able to determine the ground state. If we add the
following fine-tuned hole-related terms to the initial Hamiltonian~\eqref{Ham}
\begin{eqnarray}
  H^{\rm hole}_{\rm canonical}= - t_{\rm hole}\sum_{<ijk>}\Bigg\{\Bigg[\Big|\raisebox{-0.5ex}{\includegraphics{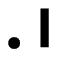}}\Big>\Big<\raisebox{-0.5ex}{\includegraphics{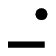}}\Big| + \mathit{h.c.}\Bigg]  \nonumber\\
  -\Big|\raisebox{-0.5ex}{\includegraphics{Ham22.eps}}\Big>\Big<\raisebox{-0.5ex}{\includegraphics{Ham22.eps}}\Big| -\Big|\raisebox{-0.5ex}{\includegraphics{Ham23.eps}}\Big>\Big<\raisebox{-0.5ex}{\includegraphics{Ham23.eps}}\Big|\Bigg\}
\label{Ham2}
\end{eqnarray}
then the resulting ground state becomes
\begin{equation}
  \ket{ G_{N_h}}=\frac{1}{\sqrt{Z(w^2)}}\sum_{\{\mathcal{C}_{N_h}\}}
  w^{\displaystyle{N_p[\mathcal{C}_{N_h}]}}\ket{\mathcal{C}_{N_h}} \;,
  \label{Gzw2}
\end{equation}
where the number of holes $N_h$ is now fixed at a specified value. The norm of this
wave-function, $Z(w^2)$,  is the \emph{canonical} partition function for the
set of dimer coverings with a fixed number of holes.

On the other hand, if we add the following terms,
which do not conserve the number of holes in the system, to the Hamiltonian~\eqref{Ham}
\begin{eqnarray}
  H^{\rm hole}_{\rm {grand-canonical}} =
  - \tilde t_{\rm hole}\sum_{links}\Bigg\{ \Bigg[\Big|\raisebox{-0.5ex}{\includegraphics[scale=0.6]{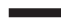}}\Big>\Big<\raisebox{-0.5ex}{\includegraphics[scale=0.6]{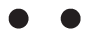}}\Big| + h.c.\Bigg]  \nonumber\\
  - z^2\Big|\raisebox{-0.5ex}{\includegraphics[scale=0.6]{Ham20.eps}}\Big>\Big<\raisebox{-0.5ex}{\includegraphics[scale=0.6]{Ham20.eps}}\Big|-z^{-2}\Big|\raisebox{-0.5ex}{\includegraphics[scale=0.6]{Ham21.eps}}\Big>\Big<\raisebox{-0.5ex}{\includegraphics[scale=0.6]{Ham21.eps}}\Big| \Bigg\}
  \nonumber \\
  &&
\label{Ham3}
\end{eqnarray}
then the ground state of the system becomes
\begin{equation}
  \ket{ G}=\frac{1}{\sqrt{Z(w^2,z^2)}}\sum_{\{\mathcal{C}\}}
  w^{\displaystyle{N_p[\mathcal{C}]}} z^{\displaystyle{N_h[\mathcal{C}]}}
  \ket{\mathcal{C}} \;,
\label{Gzw3}
\end{equation}
with
\begin{equation}
 Z(w^2,z^2)=\sum_{\{\mathcal{C}\}} w^{2N_p[\mathcal{C}]}
 z^{2N_h[\mathcal{C}]} \;.
 \label{Zzw}
\end{equation}
Equation~\eqref{Ham3} includes a natural, short-range repulsion between
neighboring holes and an off-diagonal term which represents
creation-annihilation of dimers. Furthermore, there is a dimer fugacity term,
as in every perturbative derivation of a quantum dimer model.\cite{rokhsar88}
We note that Eq.~\eqref{Zzw} has the same form as a grand partition function
for dimer coverings of the square lattice.  This partition function now
depends not only on the interaction~$u$ defined below Eq.~\eqref{z1}, but also
on the hole chemical potential $\mu/|u| = 2\ln z$.

Since the canonical and grand-canonical ensembles become equivalent in the
thermodynamic limit, the two ground-state wave functions \eqref{Gzw2}
and~\eqref{Gzw3} must correspond to the same ground-state physics.
Furthermore, it is clear that for $w=1$ the models are located at the usual RK point
of the quantum dimer model on the square lattice.  For a system with periodic
boundary conditions, each configuration $\mathcal{C}$ contains only an
\emph{even} number of holes, with half of the holes on either sublattice.

We remark that the fact that the exact ground-state wave function is a sum (as
opposed to a product) of the ground states of sectors labeled by the number of
holes on the lattice, is due to the resonance term that we have used to
represent the motion of holes. In particular, we have assumed that a dimer can
\emph{break} into two holes which themselves repel each other. In the limit of
very strong hole-hole repulsions, in strong-coupling perturbation theory, it
is straightforward to recover a fixed-hole density sector with a single-hole
resonance move involving three sites in any direction; the coupling strength
in Eq.~\eqref{Ham2} then becomes $t_{\rm hole}\sim z^4\tilde t_{\rm hole}$,
and thus, in the limit $z\rightarrow 0$, it reduces to an effective hopping amplitude for the holes.

\section{Mean-Field Results}
\label{sec:mean-field}

To examine the physics described by the ground-state wave functions
obtained in the previous section we start with a discussion of a
mean-field theory of the phase diagram.  We use the standard approach of
regarding the probability densities, obtained by squaring the wave
function, as the Gibbs weights of a classical two-dimensional system and
focus on the interacting dimer model on the square lattice at finite
hole density.  Although mean-field theory is
insufficient to describe two-dimensional critical systems, it is a
useful tool to obtain qualitative features of the phase diagram as well
as the behavior deep in the phases, away from criticality.

The details of this theory are presented in
Appendix~\ref{app:mean-field-details}. We begin by constructing a
Grassmann representation of the partition function for an interacting
dimer model using the standard methods introduced by
Samuel.\cite{samuel80,Samuel} The resulting theory involves Grassmann
(anti-commuting) variables residing on the sites of the square lattice.
The action of the Grassmann integral is non-trivial and is parametrized
by the hole fugacity $z$ and a coupling between dimers $V=z^2 (e^u-1)$,
where $u=-2\ln w$.

Since the action of the resulting Grassmann path integral is not
quadratic in the Grassmann variables, it cannot be reduced to the
computation of a determinant. Thus, we use a standard mean-field
approach which, in this case, involves the introduction of two
Hubbard-Stratonovich (bosonic) fields $\phi_i$ and $\chi_{ij}$, defined
on the sites and links of the square lattice respectively. Upon
integrating out the Grassmann variables one obtains an effective theory
for the fields $\phi_i$ and $\chi_{ij}$ which, as usual, is solved
within a saddle-point expansion. The dimer $m_0$ and hole $n$ densities,
as well as the columnar order parameter $m$, can be expressed
straightforwardly in terms of the fields $\phi_i$ and $\chi_{ij}$. From
this effective theory one can compute an effective potential $\Gamma$
and the configurations of the observables of interest, $n$, $m_0$ and
$m$, as functions of $z$ and $V$, and determine the phase diagram.

\begin{figure}
\subfigure[]{
 \includegraphics[width=0.42\textwidth]{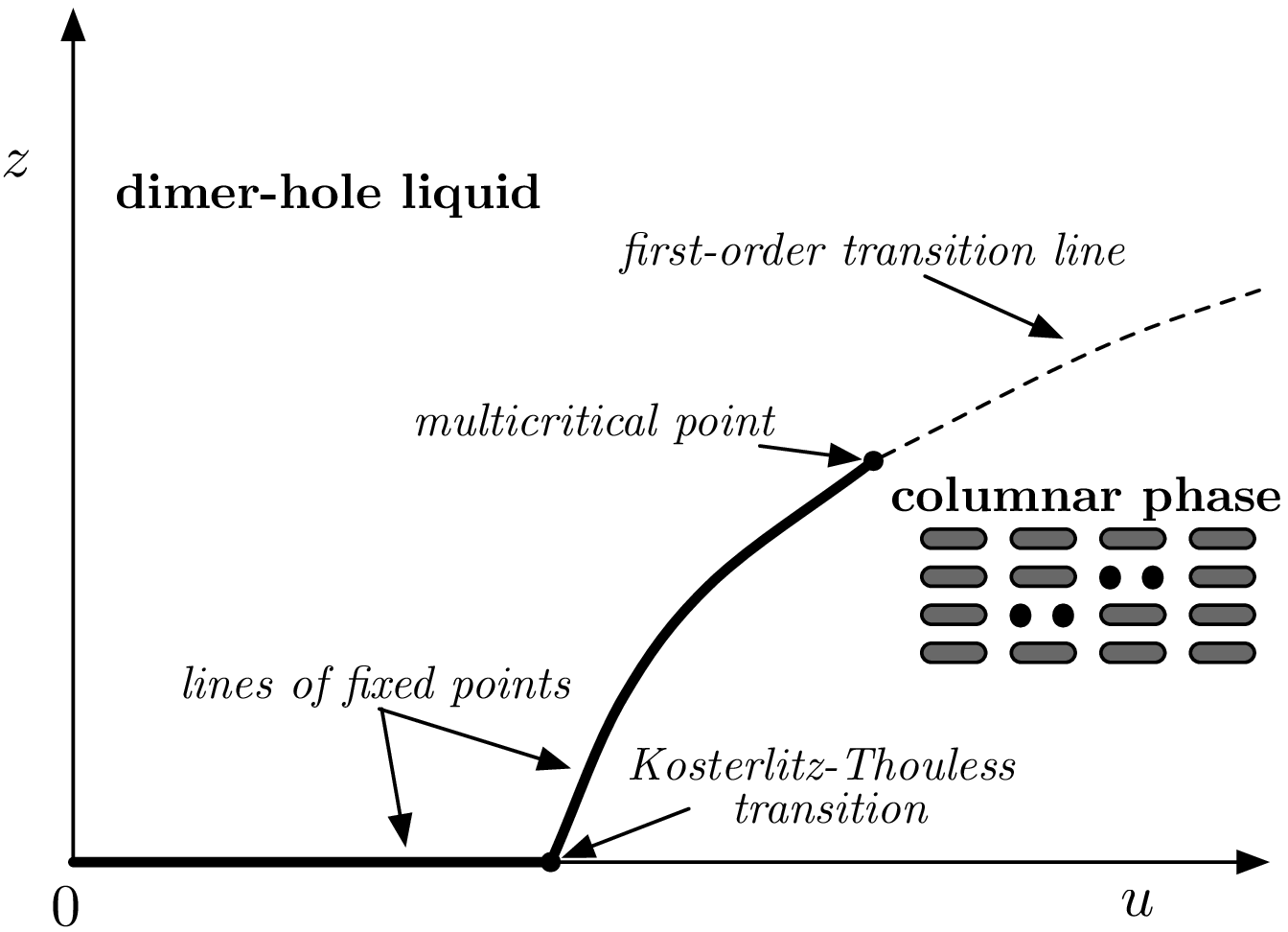}
}
\subfigure[]{
  \includegraphics[width=0.42\textwidth]{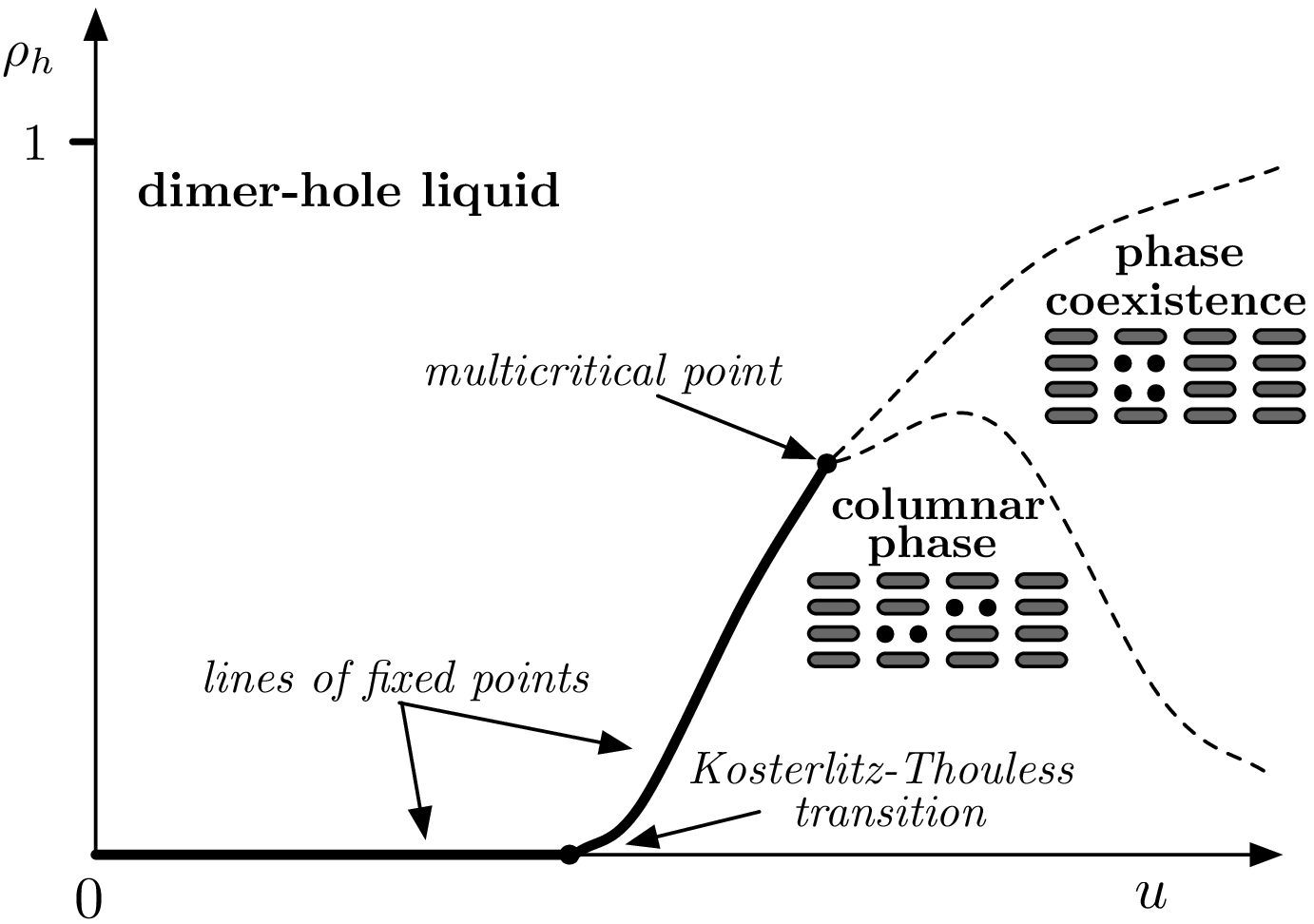}
}
\caption{Qualitative phase diagrams for the interacting doped dimer
model. a) Phase diagram as a function of the fugacity $z$ for the holes
and $u=-\ln w$, the dimer interaction coupling constant. The line of
continuously varying exponents for low fugacity evolves to a line of
first-order transitions. b) Phase diagram in terms of the hole density
$\rho_h$ and the coupling constant~$u$. Here the line of first-order
transitions opens up into a two-phase coexistence region.}
\label{fig:sketch}
\end{figure}

As a function of hole density (or hole fugacity) and $u$, we find that the
phase diagram has two phases (shown qualitatively in
Fig.~\ref{fig:sketch}), namely a dimer-hole liquid phase and a hole-poor
phase with long-range columnar order. The nature of the 
transitions between these phases is incorrectly described by the mean-field theory,
particularly at zero doping and near the tricritical point.
The correct behavior is the subject of a detailed analysis in the
subsequent sections. Nevertheless, the mean-field phase diagram
correctly predicts
that at low hole densities and moderate values of $u$ the
transition between the dimer-hole liquid and the columnar solid phase is
continuous, that for large $u$ 
the transition is first order, and that there is a tricritical point at
\begin{equation}
u_{tr} \simeq 2.733, \qquad z_{tr} \simeq 0.075 \;.
\end{equation}
Remarkably, the Monte Carlo simulations presented in
Section~\ref{sec:MC} yield a tricritical point at a location quite
consistent with these values.

Another correct prediction of the mean-field theory is the behavior of
the connected hole density correlation function deep in the dimer-hole
liquid phase. This prediction, also discussed in detail in Appendix
\ref{app:mean-field-details}, fits surprisingly well the Monte Carlo
simulations of Krauth and Moessner,\cite{krauth03} performed for a
system of classical non-interacting ($u=0$) dimers at finite hole
density. The mean-field result is consistent with the simulations for a
quite broad region of densities even quite close to the fully packed
limit $z \to 0$ where, naturally, a wrong correlation length exponent is
predicted.

\section{Phase Diagram and Correlations for Interacting Dimers and
Holes}
\label{sec:exact-critical-behavior}

We now turn to a more precise analysis of the phase transitions of the
interacting classical dimer models as a function of hole density on
interaction parameter $u$. Here we take advantage of a wealth of
information and methods from two-dimensional systems, exact solutions
and conformal field theory, to analyse the behavior in detail and
extract conclusions that will be quite useful for the analysis of the
wave functions. We begin with a discussion of the undoped case, and then
discuss the physics at finite hole density.

\subsection{Interacting Dimers at Zero Hole Density}
\label{sec:zero-density}

It is a well known fact that both classical and quantum two-dimensional
dimer models can be represented in terms of height
models. For the classical case, this mapping is well
known\cite{blote82a,blote82b,henley97a,ardonne04}. The mapping for the
quantum case has also been discussed
extensively\cite{fradkin90a,levitov90,zheng-sachdev89,fradkin91,ardonne04,fradkin04}.
In both cases the mapping relates each dimer configuration to a
configuration of a set of integer-valued (height) variables, $h(\vec
r)$, which reside on the sites of the dual lattice, a square lattice in
the case of interest here. Thus, this mapping amounts to a duality
transformation.
 
An alternative picture follows from realizing that dimer configurations
can be mapped onto the degenerate ground-state configurations of fully
frustrated Ising models\cite{blote82a,moessner03a}. In our case,
 the corresponding spin model is the fully frustrated Ising
model on the square lattice (FFSI) at zero temperature. In the Ising
model picture, each dimer is dual to an unsatisfied bond of the fully
frustrated Ising model and classical dimer interactions
correspond to second neighbor interactions in the square lattice
FFSI\cite{fradkin81}. It is easy to see that holes correspond to
unfrustrated plaquettes in the FFSI. In this work we use primarily
the language of the height representation.
 
To map the square-lattice classical dimer model onto a height model, we
follow the prescription given in Ref.~\onlinecite{ardonne04}.  One first
assigns a height variable to each plaquette.  In going around a vertex
on the even sub-lattice clockwise, the height changes by $+3$ if a dimer
is present on the link between the plaquettes, and by $-1$ if no dimer
is present on that link. On the odd sub-lattice, the heights change by
$-3$ and $+1$ respectively. The dimer constraint, that one lattice site
belongs to one and only one dimer, implies (for the square lattice)
that, for the fully packed case, there are four possible configurations
of dimers for each lattice site. In the dual height model this is
reproduced by the period four property $ h \equiv h+4$ of the allowed
height configurations. It is easy to see that for the allowed
configuration the average values that the height variables can take at a
given site of the direct lattice (a vertex) are $\pm 3/2, \pm 1/2$. On
the other hand, a uniform shift {\em of all} the heights by one unit, $h
\to h+1$, leads to an equivalent state. This mapping works strictly
speaking only for the fully packed case. Holes are sites that don't
belong to any dimer and thus represent violations of the full packing
rule. They play the role of topological defects (``vortices'') in the
(dual) height representation.

The exact solution of the non-interacting fully packed dimer model on
the square (and other) lattice has been known for a long
time\cite{fisher63}. In particular the long-distance behavior of the
dimer density correlation functions and the hole density correlation
functions are known explicitly\cite{fisher63,youngblood80}. These
correlation functions obey power law behaviors and show that this is a
critical system. Here we will use the standard approach to map the exact
long distance behavior of two-dimensional critical systems to the
behavior of the simplest critical system, the Gaussian or free boson
model\cite{kadanoff77}. This approach is consistent for the free dimer
model on an even lattice with periodic boundary conditions since its
central charge (or conformal anomaly) is also $c=1$.

We will use as reference states (``ideal states'' in the terminology of
Kondev and Henley\cite{kondev-henley96}) the four columnar states of the
dimer coverings, which have the largest number of flippable plaquettes,
and use them to define an effective field theory for this
problem\cite{fradkin04}. We will assign a uniform value to the
coarse-grained height field $h=0,1,2,3$ to each reference (columnar)
state.  Let $n_x(\vec r)$ and $n_y(\vec y)$ represent the coarse grained
dimer densities of the horizontal link with endpoints at the pairs of
lattice sites $\vec r$ and $\vec r + \vec e_x$, and vertical links with
endpoints $\vec r$ and $\vec r + \vec e_y$ respectively. Here $\vec e_x$
and $\vec e_y$ are two lattice unit vectors along the $x$ and $y$
directions respectively, with a lattice spacing of $1$.

We can now define the columnar local order parameter as the two
component vector
\begin{eqnarray}
\mathcal{O}_x(x,y)&=&n_x(x,y)-n_x(x+1,y)\nonumber \\
\mathcal{O}_y(x,y)&=&n_y(x,y)-n_y(x,y+1)
\end{eqnarray}
which clearly correspond to the $(\pi,0)$ and $(0,\pi)$ Fourier
components of the dimer densities. This two-component order parameter
takes four distinct values for each one of the columnar states, and
changes sign under shifts by one lattice spacing in either direction. It
also transforms as a vector under $90^{\circ}$ rotations. Thus it is the
order parameter for columnar order.

\subsubsection{Effective Field Theory: the non-interacting case}
\label{sec:dimer-qft}

The fluctuations of the free field $h(\vec r)$ are described by a
continuum Gaussian (free boson) model. We will find it simpler to work
with the rescaled height field $\phi=\frac{\pi}{2} h$. For this field
the periodicity condition $h \to h+4$ becomes $ \phi \to \phi + 2 \pi$.
(For the rescaled field the ideal states are $\phi=0,\pi/2,\pi,3\pi/2$.)
Thus, the allowed operators are $2\pi$ periodic functions of $\phi$, and
are either derivatives of $\phi$, or the exponential (or {\em charge})
operators $\exp(\pm i \phi)$, $\exp(\pm 2 i \phi)$, $\exp(3 i \phi)$ and
$\exp(\pm 4 i \phi)$, which are $2\pi$ periodic functions of $\phi$.

The action $S$ for the rescaled field is
\begin{equation}
S=\int d^2x \frac{K}{2} \left( \nabla \phi\right)^2
\label{eq:gaussian-action}
\end{equation}
For the free dimer model the stiffness is $K=\frac{1}{4\pi}$
(see below).

By matching the exact correlation functions of the free dimer model on
the square lattice one readily finds the following operator
identification of the coarse-grained dimer densities in terms of free
field operators\cite{fradkin04}
 \begin{eqnarray}
n_x-\frac{1}{4}&=&\frac{1}{2\pi}(-1)^{x+y}\partial_y \phi +
\frac{1}{2}[(-1)^{x} e^{\displaystyle{i \phi }}+ {\rm c.c.}]\\
n_y-\frac{1}{4}&=& \frac{1}{2\pi}(-1)^{x+y+1}\partial_x \phi+
\frac{1}{2}[(-1)^{y} i \; e^{\displaystyle{i \phi}}+ {\rm c.c.}]\nonumber
 \\
 &&
\label{densities}
\end{eqnarray}
In Ref.~\onlinecite{fradkin04} it was shown that this is an operator
identity for the free dimer model on the square lattice in the sense the
the asymptotic long-distance behavior of the dimer density correlation
functions computed with this Gaussian model are the same as the exact
long distance correlation functions for the free dimer problem on the
square lattice\cite{fisher63,youngblood80} provided the stiffness
$K=\frac{1}{4\pi}$. Notice that, with this identification, when the
field $\phi$ takes each of the values $0,\pi/2,\pi,3\pi/2$ (the ``ideal
states'') the density operators take four distinct values which reflect
the broken symmetries of the four columnar states.

From the operator identification of Eq.~\eqref{densities} 
the columnar order parameter is, up to a 
normalization constant, 
\begin{equation}
\mathcal{O}_x  = \cos \phi , \qquad
\mathcal{O}_y = \sin\phi
\end{equation}
Due to the effects of dimer-dimer interactions the form of this effective action is
\begin{equation}
S=\int d^2 x\; \frac{K}{2} \left(\nabla \phi\right)^2+\textrm{perturbations}
\end{equation}
The effect of the interactions is a finite renormalization of the
stiffness $K$ away from its free dimer value, $K=\frac{1}{4\pi}$.

We saw above that, due to the dimer constraints, the allowed charge
operators are $O_n(\vec r)=\exp(i n \phi(\vec r))$. We also saw that the
columnar order parameter is proportional to the operator
$\frac{1}{2}(O_1(\vec r)+ O_{-1})(\vec r)=\cos \phi(\vec r)$, and
carries the unit of charge $n=1$. One can also define {\em vortex} or
{\em magnetic} operators\cite{kadanoff77}, and example of which is the
hole. A vortex operator causes the field $\phi$ to wind by $2\pi m$,
where $m$ is the vorticity (or {\em magnetic charge}).
One can similarly define a general composite operator $O_{n,m}(\vec r)$
with $n$ units of (electric) charge and $m$ units of vorticity (or
magnetic charge).
Its scaling dimensions, $\Delta(n,m)$
are\cite{kadanoff77}
\begin{equation}
\Delta_{n,m}(K)=\frac{n^2}{4\pi K}+\pi K m^2
\end{equation}

We can now use these results to identify a few operators of interest and
give their scaling dimensions. These results are summarized in Table
\ref{table:0density}
\begin{enumerate}
\item 
The columnar order parameter is the elementary charge operator
  $O_{\pm 1,0}$ and has no vorticity. On the (columnar) ideal states
  $0$, $\pi/2$, $\pi$, and $3\pi/2$ this operator takes the values $1$
  ,$i$ ,$-1$, and $-i$ respectively. Its scaling dimension is
  $\Delta_{1,0}(K)=\frac{1}{4\pi K}$. At the free dimer point,
  $K=\frac{1}{4\pi}$, its scaling dimension is
  $\Delta_{1,0}(\frac{1}{4\pi})=1$. This is consistent with the exact
  results\cite{fisher63,youngblood80} that the density correlation
  function falls off as $1/r^2$. The operator identification of
  Eq.~\eqref{densities} is based on these facts\cite{fradkin04}.
\item 
The operator $O_{\pm 2,0}=\exp(\pm 2i \phi)$ takes the values $1$,
  $-1$, $1$ and $-1$ on each of the ideal columnar states. It is clearly
  the order parameter for symmetry breaking by $90^\circ$ rotations: it
  is the order parameter for {\em orientational} symmetry.
\item 
The operator with the highest allowed electric charge is
  $O_{\pm4,0}=\exp(\pm4i\phi)$. Its scaling dimension is
  $\Delta_{4,0}(K)=\frac{4}{\pi K}$. At the free dimer point it has
  dimension $\Delta_{4,0}(\frac{1}{4\pi})=16$, and it is a strongly
  irrelevant operator.
  This operator arises naturally due to the fact that the microscopic
  heights $h$ take integer values, and hence height configurations which
  differ by an uniform integer shift are physically equivalent. This
  operator does not break any physical symmetry of the dimer model.
\item 
The hole operator is represented by the fundamental vortex
  operator $O_{0,\pm 1}$. A vortex with unit positive magnetic charge
  corresponds to a hole on one sublattice, and a vortex with unit
  negative magnetic charge to a hole on the other sublattice. The
  scaling dimension of the vortex (hole) operator is
  $\Delta_{0,1}(K)=\pi K$. At the free dimer value, the scaling
  dimension of the hole operator is
  $\Delta_{0,1}(\frac{1}{4\pi})=\frac{1}{4}$, which is consistent with
  the exact result that the hole-hole correlation function decays
  $1/\sqrt{r}$ at large distances\cite{fisher63}.
\item 
The operator which describes a pair of holes on nearby sites of
  the {\em same} sublattice is represented by the operators $O_{0,\pm
  2}$ which carry two units of magnetic charge (vorticity). This
  operator creates (or destroys) a {\em
  diagonal dimer} connecting nearby points on the same
  sublattice.\cite{read-sachdev91} In the $2+1$-dimensional quantum
  dimer model, this operator is useful to describe the possible pairing
  of holes. This operator has dimension $\Delta_{0,2}(K)=4\pi K$. Its
  scaling dimension at the free dimer point is
  $\Delta_{0,2}(\frac{1}{4\pi})=1$, and is relevant for $K<\frac{1}{2\pi}$.  As noted in Refs. \onlinecite{fendley02} and  \onlinecite{ardonne04}, this operator maps the square lattice into a deformed triangular lattice. The irrelevancy of this operator for $K>\frac{1}{2\pi}$ implies that this line of fixed points also exists for a deformed triangular lattice, as discussed recently in Ref. \onlinecite{trousselet07}.
\item 
The free dimer problem is a free fermion system, which can be
  solved by Pfaffian methods.\cite{fisher63,samuel80,fendley02} This is
  actually a theory with two free real (Majorana) fermions or,
  equivalently, one free complex (Dirac) fermion, at its (massless)
  fixed point, whose central charge is also $c=1$. The appropriate
  fermion operator is a composite operator of the order and disorder
  operators\cite{kadanoff77,kadanoff-ceva71} which, in this case, is
  $O_{\frac{1}{2},1}$. At the free dimer point the fermion operator has
  scaling dimension $\Delta_{\frac{1}{2},1}(\frac{1}{4\pi})=\frac{1}{2}$
  and (conformal) spin $nm=\frac{1}{2}$ (as it should for a free
  fermion).  At particular values of $K$ it is also possible to define
  parafermion operators\cite{fradkin-kadanoff80,zamolodchikov-fateev85}
  operators which obey fractional statistics.
  For instance, at the KT point, $K=\frac{2}{\pi}$ (see below), the
  operator $O_{1,\frac{1}{4}}$ has dimension $\frac{1}{4}$ and spin
  $\frac{1}{4}$, and it is a {\em semion}. In fact, and not surprisingly, a fermionic approach\cite{fendley02} can be used to map this critical line onto an Euclidean version of the Luttinger-Thirring model. The coarse-grained height model description we sue here corresponds to the bosonization approach of the fermionic version of this problem.
\end{enumerate}

\begin{table}[t]
\newcolumntype{Y}{>{\centering\arraybackslash$}m{1.0cm}<{$}}
\newcolumntype{C}{>{\centering\arraybackslash$}m{1.30cm}<{$}}
\renewcommand{\arraystretch}{2}
\begin{tabular}{|C||C|C|C|C|C|C|}
\hline
&{\rm columnar}&{\rm rotational}&&{\rm hole}&{\rm hole\; pair}\\
  	&O_{1,0} & O_{2,0}  & O_{4,0} & 
	O_{0,1}& O_{0,2}\\
\hline
 K=\frac{1}{4\pi}& 1 & 4 & 16 & \frac{1}{4} & 1\\
\hline
 K=\frac{2}{\pi}& \frac{1}{8} & \frac{1}{2} & 2 & 2 & 8\\
\hline
\end{tabular}
\label{table:qn}
\caption{Scaling dimensions of the order parameters (charge) and hole (vortex) operators at the free dimer point, $K=\frac{1}{4\pi}$, and at the KT point, $K=\frac{2}{\pi}$. }
\label{table:0density}
\end{table} 

\subsubsection{Effective Field Theory: interactions and phase transition}
\label{sec:KT}

We now turn to the effects of dimer-dimer interactions. We recall that
we are considering only interactions of a pair of dimers in a plaquette.
The interaction energy is
\begin{equation}
H_{\rm int}=-u \sum_{\vec r} \Big[ n_x(\vec r) n_x(\vec r+\vec e_y)+n_y(\vec r) n_x(\vec r+\vec e_x)\Big]
\end{equation}
We wish to find the corresponding operator in terms of the coarse
grained (rescaled) height field $\phi$. This can be done by using the
operator product expansion (OPE) of the coarse-grained form of the dimer
density operators, Eq.~\eqref{densities}, in terms of the field $\phi$.
We will also need the standard OPE of the (normal ordered) charge
operators\cite{witten78,kadanoff-brown79,boyanovsky89a}
\begin{eqnarray}
&&:\cos (n \phi(x)) : \; :\cos (n \phi(y)):= \frac{1}{2} :\cos (2n\phi(x)):\nonumber \\
&&+\frac{1}{(\mu^2|x-y|^2)^{n^2}}\Big[1-n^2|x-y|^2 :\left(\nabla \phi \right)^2:+\ldots\Big] \nonumber \\
&& \\
&&:\sin (n \phi(x)) : \; :\sin (n  \phi(y)):= - \frac{1}{2} :\cos (2n\phi(x)):\nonumber \\
&&+\frac{1}{(\mu^2|x-y|^2)^{n^2}}\Big[1-n^2 |x-y|^2 : \left(\nabla \phi \right)^2:+\ldots \Big]\nonumber \\
&&
\label{ope1}
\end{eqnarray}
where $\mu$ is a short-distance cutoff and the ellipsis represents the
contributions of irrelevant operators. Using these results we find that
the net effect of the interactions is to renormalize the stiffness $K$
upwards
\begin{equation}
K=\frac{1}{4\pi}+\frac{1}{2}\left(1+\frac{1}{\pi^2}\right) u+ \mathcal{O}(u^2)
\label{Kint}
\end{equation}
where we have denoted $u>0$ for {\em attractive} interactions. This
expression for the renormalization of $K$ is only accurate to linear
order in the dimer-dimer interaction. Higher order renormalizations (in
$u$) would result if the effects of irrelevant operators are also taken
into account. The relation between $K$ and the microscopic model is
non-universal and can only be determined either from an exact solution
or from a numerical simulation. One can determine the function $K(u)$
from the Monte Carlo simulations we present elsewhere in this paper.
What is important is that these non-universal effects affect only the
relation between the coefficients of the effective theory and not the
form of the effective theory itself. Thus, the effective action of the
field theory for the interacting classical dimer model at zero hole
density has the form
\begin{equation}
S=\int d^2x \left[ \frac{K}{2} \left(\nabla \phi\right)^2+g \cos(4\phi)\right]
\label{Seff-0density}
\end{equation}
where we have included the effects of the charge $4$ perturbation, $\cos
(4\phi)=\cos(2\pi h)$, which 
biases the coarse-grained height field to take integer values.

For an {\em anisotropic} dimer-dimer interaction, which arises form a
term which weights differently the interactions between parallel
horizontal dimers from those of parallel vertical dimers, we would have
also found a $\cos(2\phi)$ operator in
addition to an anisotropy for the stiffness. Thus, an anisotropy in the
dimer-dimer interaction is a relevant perturbation which couples to the
{\em orientational} order parameter $\cos(2\phi)$.

In Table \ref{table:0density} we see that as the attractive interactions
grow there will be a critical value of the interaction $u$ at which the
stiffness $K(u_c)=\frac{2}{\pi}$. At this point, the $\cos(4\phi)$
operator has scaling dimension $\Delta_{2,0}=2$, where it becomes
marginal. For $u>u_c$ ($K>K(u_c)$) this operator becomes relevant. We
also see that at $K(u_c)=\frac{2}{\pi}$ the columnar order parameter,
$\cos \phi$, has scaling dimension $1/8$ and it is the most relevant
operator in this problem. Thus this is a phase transition from a {\em
critical phase}, for $K<\frac{2}{\pi}$, without long-range order but
with power law correlations, to a phase with long-range columnar order,
for $K>\frac{2}{\pi}$, in which the columnar order parameter has a
non-vanishing expectation value. This is a standard Kosterlitz-Thouless
(KT) transition\cite{kosterlitz73,kosterlitz74,jose77} which is
naturally described by the sine-Gordon field
theory\cite{luther75,wiegmann78,amit80,dennijs81} whose (Euclidean)
action is given in Eq.~\eqref{Seff-0density}. The only difference between
this problem and the standard KT transition, {\em e.g.\/} the classical
2D XY model, is that the phase with a finite correlation length is
ordered: it is a columnar state with a four-fold degenerate non-uniform
state, whereas the finite-correlation length phase of the XY model (and
of its dual surface roughening model) has a non-degenerate translation
invariant state. In spite of these global differences, this phase
transition is in the KT universality class. Thus the well known behavior
of the correlation functions at the KT transition apply to this case as
well \cite{alet05,alet06,Castelnovo06,Poilblanc06}. In Section
\ref{sec:MC} we verify this behavior by a detailed Monte Carlo study of
the columnar and orientational order parameters and of their associated
susceptibilities for the interacting dimer model.

\subsection{Interacting Dimers at Finite Hole Density}
\label{sec:finite-density}

We now consider the dimer model at finite hole density, away from the
full packing condition. The classical partition function for this
problem is given in Eq.~\eqref{Zzw}, where the weights (fugacities) $z$
and $w$ are related to the coupling constant $u$ and the chemical
potential $\mu$ as described earlier. Recall that the 2D classical
partition function $Z(w,z)$ is the norm of the ground-state wave
function $\ket{G}$, of Eq.~\eqref{Gzw}, of the 2D doped quantum dimer
model. In terms of a sum over configurations of electric charges $n$ and
magnetic charges (vortices) $m$, the partition function $Z(w,z)$ is
equivalent to that of a generalized (neutral) Coulomb gas (GCG) of
electric and magnetic charges in two
dimensions\cite{kadanoff78,nienhuis87}
\begin{eqnarray}
Z(w,z)=\!\!\!\!\!\!{\sum_{\{ n(\vec r), m(\vec R)\} }}^{\!\!\!\!\!\!\!\! \prime} \;
\exp\Bigg[
\frac{N^2}{4\pi K} \sum_{\vec r,\vec r^{\; \prime}} n(\vec r) \; \ln |\vec r-\vec r^{\;\prime}|\; n(\vec r^{\; \prime})\nonumber\\
+\pi K \sum_{\vec R,\vec R^{\; \prime}} m(\vec R) \; \ln |\vec R-\vec R^{\;\prime}| \; m(\vec R^{\; \prime})\Bigg]
\nonumber \\
\times\; \exp\Bigg[\sum_{\vec r}\ln w \;  n(\vec r)^2+\sum_{\vec R} \ln z\; m(\vec R)^2 \nonumber\\
-i \sum_{\vec r, \vec R} N \; n(\vec r) \; m(\vec R) \; \Theta(\vec r-\vec R)\Bigg]\;\;\;\;\;
\label{GCG}
\end{eqnarray}
where prime denotes that the sum is restricted to neutral configurations
with vanishing total charge and vanishing total vorticity, {\it i.e.\/}
$\sum_{\vec r} n(\vec r)=\sum_{\vec R} m(\vec R)=0$.  Here $\Theta(\vec
r - \vec R)$ is the angle between a vortex at $\vec R$, as seen from a
charge at $\vec r$, measured with respect to the (arbitrary) $x$ axis.
It is the Cauchy-Riemann dual of the logarithm,
\begin{eqnarray}
G(\vec r -\vec r^{\;\prime})&=&-\ln |\vec r-\vec r^{\;\prime}| 
 \\
\Theta(\vec r-\vec r^{\;\prime})&=&-\tan^{-1}\left(\frac{y-y^\prime}{x-x^\prime}\right) \\
-\nabla^2 G(\vec r - \vec r^{\;\prime})&=&2\pi \; \delta^2(\vec r - \vec r^{\;\prime})\\
\partial_\mu G&=&\epsilon_{\mu \nu} \partial_\nu \Theta
\end{eqnarray}
As usual, the logarithmic interaction is regularized so that it vanishes
for $\vec r=\vec r^{\prime}$. The short distance behavior of the
interactions is absorbed in the fugacities $z$ and $w$.

For the case we are discussing here, the GCG of interest has $N=4$, and
Eq.~\eqref{GCG} is just the the Coulomb-gas form of the partition
function of the $\mathbb{Z}_4$ model, and for the related 2D
Ashkin-Teller model. This is a well understood system \cite{jose77,wiegmann78,boyanovsky89a,lecheminant02} and in our
case, it corresponds to the grand-partition function for the doped
interacting dimer model on the square lattice at low hole
densities.

In the limit of low fugacities, $z\ll 1$ and $w\ll 1$, the GCG is
equivalent to a (generalized) sine-Gordon field theory in
two-dimensional Euclidean space-time, whose (Euclidean) Lagrangian is
given by\cite{wiegmann78,boyanovsky89a}
\begin{equation}
\mathcal{L}_E=\frac{1}{2} \left(\partial_\mu \phi\right)^2-\frac{2z}{a^2} \cos\left(\frac{N}{\sqrt{K}}\; \phi\right)-\frac{2w}{a^2} \cos \left(2\pi \sqrt{K} \; \widetilde \phi \right)
\label{eff-action-euclidean}
\end{equation}
Here $\widetilde \phi$ is the dual field, defined by
\begin{equation}
\epsilon_{\mu \nu} \partial_\nu \phi=i \partial_\mu \widetilde \phi
\end{equation}
In Eq.~\eqref{eff-action-euclidean} we can see by inspection that the
operators $\cos(\frac{N}{\sqrt{K}}\; \phi) $ and $\cos(2\pi\sqrt{K}\;
\widetilde \phi)$ can be identified respectively with the operators
$\frac{1}{2}(O_{N,0}+O_{-N,0})$ and $\frac{1}{2}(O_{0,1}+O_{0,-1})$
discussed above.

We will use this effective field theory to study the transition between
the liquid and the ordered phases of the interacting dimer model. At
$z=w=0$ this is the KT transition discussed above. For general $N$, both
operators have the scaling dimension if $\frac{N}{\sqrt{K}}=2\pi
\sqrt{K}$, {\it i.e.\/} the theory is self-dual, which happens for
$K=\frac{N}{2\pi}$. For $N=4$, $K=\frac{2}{\pi}$, both operators have
scaling dimension $2$ and both are marginal. This is the only case we
will discuss here. (A detailed discussion of the more general case of
$N>4$ was given by Lecheminant {\it et al\/}.\cite{lecheminant02})

For $N=4$ the Euclidean Lagrangian becomes\cite{lecheminant02}
\begin{equation}
\mathcal{L}_E=\frac{1}{2} \left(\partial_\mu \phi\right)^2-\frac{2z}{a^2} \cos\left(\sqrt{8\pi } \; \phi\right) -\frac{2w}{a^2} \cos\left(\sqrt{8\pi} \; \widetilde \phi \right)
\label{self-dual}
\end{equation}
It turns out that this a problem which can be solved
exactly.\cite{ginsparg88,lecheminant02} The most direct way of doing
this is to perform an analytic continuation from 2D Euclidean space to
$1+1$-dimensional Minkowski space-time, {\it i.e.\/} to think of this
problem as a $1+1$-dimensional quantum field theory. The {\em
Hamiltonian density} of the equivalent $1+1$-dimensional field theory is
\begin{eqnarray}
\mathcal{H}&=&\frac{1}{2}(\partial_x \widetilde \phi)^2+\frac{1}{2}(\partial_x \phi)^2 \nonumber\\
&&-\frac{2z}{a^2}\cos\left(\sqrt{8\pi}\;\phi\right)-\frac{2w}{a^2}\cos\left(\sqrt{8\pi}\;\widetilde \phi \right)\nonumber \\
 &=& \frac{1}{2}(\partial_x \widetilde \phi)^2+\frac{1}{2}(\partial_x \phi)^2  \nonumber\\
&&-2\frac{(z+w)}{a^2} \cos (\sqrt{8\pi} \phi_L)\; \cos (\sqrt{8\pi} \phi_L)
\nonumber \\ 
&&+2\frac{(z-w)}{a^2} \sin (\sqrt{8\pi} \phi_L)\; \sin (\sqrt{8\pi} \phi_L)
\label{1+1d-H}
\end{eqnarray}
where we have used the fact that $\Pi$, the canonical momentum conjugate
to the field $\phi$ is simply related to the dual field $\widetilde
\phi$,
\begin{equation}
\Pi=\frac{\delta \mathcal{L}_M}{\delta \phi}=\partial_t \phi=-\partial_x \widetilde \phi
\label{Pi-dual}
\end{equation}
and obey equal-time canonical commutation relations
\begin{equation}
\left[\phi(x),\Pi(y)\right]=i \delta(x-y)
\label{ccr}
\end{equation}
In Eq.~\eqref{1+1d-H} we used the decomposition of the field $\phi$ and
the dual field $\widetilde \phi$ into right and left moving fields
$\phi_R$ and $\phi_L$,
\begin{eqnarray}
\phi&=&\phi_L+\phi_R \qquad \widetilde \phi=\phi_L-\phi_R 
\nonumber \\
\phi_L&=&\frac{1}{2}(\phi+\widetilde \phi ) \qquad \phi_R=\frac{1}{2} (\phi - \widetilde \phi )
\label{right-left}
\end{eqnarray} 
whose propagators are
\begin{eqnarray}
\langle \phi_L(z) \phi_L(w) \rangle=-\frac{1}{4\pi} \ln(z-w)\nonumber\\
\langle \phi_R(\bar z) \phi_R(\bar w) \rangle=-\frac{1}{4\pi} \ln(\bar z-\bar w)
\label{propagators-LR}
\end{eqnarray}
where we have used the complex coordinates (not to be confused with the coupling constants!) $z=\tau+ix=i(t+x)$ and $\bar z=\tau-ix=i(t-x)$, in imaginary and real time respectively.

It is easy to see\cite{ginsparg88,yellow,lecheminant02} that the
following dimension $1$ chiral operators
\begin{align}
&J_L^z=\frac{i}{\sqrt{2\pi}} \partial_z \phi_L  &&J_L^\pm=\frac{1}{2\pi}: e^{\displaystyle{\pm i\sqrt{8\pi} \phi_L}}: 
\nonumber\\
&J_R^z=\frac{-i}{\sqrt{2\pi}} \partial_{\bar z} \phi_R &&J_R^\pm=\frac{1}{2\pi}e^{\displaystyle{\mp i\sqrt{8\pi} \phi_R}}
\label{su2currents}
\end{align}
with $J_{L,R}^\pm=J_{L,R}^x\pm i J_{L,R}^y$, are the generators of an $su(2)_1$ Kac-Moody algebra given by the OPE\cite{yellow,lecheminant02}
\begin{eqnarray}
J_L^a(z) J_L^b(w)&=& \frac{\delta_{ab}}{8\pi^2(z-w)^2} +i \frac{\epsilon_{abc}}{8\pi^2(z-w)} J_L^c(w)\nonumber \\
J_R^a(\bar z) J_R^b(\bar w)&=& \frac{\delta_{ab}}{8\pi^2(\bar z-\bar w)^2} +i \frac{\epsilon_{abc}}{8\pi^2(\bar z-\bar w)} J_R^c(\bar w)\nonumber \\
&&
\label{su1}
\end{eqnarray}
where $a,b,c=x,y,z$ and $\epsilon_{abc}$ is the Levi-Civita tensor. We
also note for future use the following operator identifications (still
for $K=2/\pi$)
\begin{eqnarray}
e^{\displaystyle{i\sqrt{8\pi}\phi}} &\equiv& O_{4,0} 
\nonumber \\
  e^{\displaystyle{i\sqrt{8\pi} \widetilde \phi}}&\equiv&O_{0,1} 
 \nonumber \\
J_L^\pm \sim e^{\displaystyle{i\sqrt{8\pi}\phi_L}}&\equiv& O_{\pm 2,\pm \frac{1}{2}} 
\nonumber \\
  J_R^\pm \sim e^{\displaystyle{i\sqrt{8\pi} \phi_L}}&\equiv&O_{\mp 2,\pm \frac{1}{2}} 
\label{op-id1}
\end{eqnarray}

The Hamiltonian of Eq.~\eqref{1+1d-H} can be written in terms
of these generators and has the Sugawara form\cite{yellow}:
\begin{eqnarray}
\mathcal{H}&=&\frac{2\pi}{3}\left(\vec J_L \cdot \vec J_L+\vec J_R \cdot \vec J_R\right)\nonumber\\
&&-8\pi^2(z+w)J_L^x J_R^x -8\pi^2(z-w) J_L^y J_R^y
\label{1+1d-Hsu2}
\end{eqnarray}
Thus, one recovers the old result\cite{luther75,ginsparg88} that at the
KT transition the system has an effective ({\em dynamical}) $SU(2)$
symmetry.

Given the existence of an $su(2)_1$ symmetry one expects to find, in
addition to the ``spin one''(vector) representation (the $su(2)$
currents), operators associated with the spin $1/2$ representation of
$su(2)_1$. In the Wess-Zumino-Witten (WZW) version of this
theory\cite{witten84,yellow}, there is an operator with this property,
the field $g(z,\bar z)$ of the WZW model. This is a $2 \times 2$
matrix-valued field with scaling dimension $1/2$. In our theory we also
have an operator with scaling dimension $1/2$, the operator $O_{2,0}
\sim \exp(i \sqrt{2\pi} \phi)$ which we saw was related to the order
parameter for broken rotational symmetry; see Table
\ref{table:0density}. Thus, the operators of the spin $1/2$
(spinor)representation of $su(2)_1$ are identified with the operators
\begin{equation}
g(z,\bar z) \sim
\begin{pmatrix}
e^{\displaystyle{-i \sqrt{2\pi} \phi}} & e^{\displaystyle{-i \sqrt{2\pi} \widetilde \phi }} \\
e^{\displaystyle{ i \sqrt{2\pi} \phi}}  & e^{\displaystyle{ i \sqrt{2\pi} \widetilde \phi }} 
\end{pmatrix}
\sim
\begin{pmatrix}
O_{-2,0} & O_{0,-\frac{1}{2} } \\
O_{2,0}  & O_{0, \frac{1}{2} }
\end{pmatrix}
\label{spin1/2}
\end{equation}

Following the approach of Refs.~\onlinecite{ginsparg88,lecheminant02} we
will perform a global $SU(2)$ rotation by $\pi/2$ about the $y$ axis
which maps $J_{L,R}^x \to J_{L,R}^z$, $J_{L,R}^z \to -J_{L,R}^x$ and
$J_{L,R}^y \to J_{L,R}^y$, after which the Hamiltonian becomes
\begin{eqnarray}
\mathcal{H}&=&\frac{2\pi}{3}\left(\vec J_L \cdot \vec J_L+\vec J_R \cdot \vec J_R\right)\nonumber\\
&&-8\pi^2(z+w)J_L^z J_R^z -8\pi^2(z-w) J_L^y J_R^y
\label{1+1d-Hsu2-rotated}
\end{eqnarray}

Once again, one can introduce a new bose field, which we will call
$\Phi$, and its dual $\widetilde \Phi$, to use the representation of the
$su(2)_1$ current algebra of the form of Eq.~\eqref{su2currents}. Using
that $\partial_z \Phi_L=-i \partial_x \Phi_L$ and $\partial_{\bar z}
\Phi_R=i \partial_x \Phi_R$, we can rewrite the Hamiltonian as
\begin{eqnarray}
\mathcal{H}&=&\mathcal{H}_0+\mathcal{H}_{\rm pert}
\nonumber \\
\mathcal{H}_0&=&\frac{1}{2} (\partial_x \Phi)^2+\frac{1}{2} (\partial_x \widetilde \Phi)^2-
4\pi(z+w) \partial_x \Phi_L \partial_x \Phi_R
\nonumber \\
\mathcal{H}_{\rm pert}&=&-\frac{2(z-w)}{a^2} \sin(\sqrt{8\pi}\Phi_L) \sin(\sqrt{8\pi}\Phi_R)
\label{Hrotated}
\end{eqnarray}
Thus, along the phase boundary line $z=w$ the term $\mathcal{H}_{\rm
pert}$ is absent and we see that the effective Hamiltonian
$\mathcal{H}_0$ involves only marginal operators.\cite{jose77,wiegmann78,ginsparg88,lecheminant02}

Similarly, the operators in the spin-$1/2$ representation transform as
an $su(2)$ spinor under a $\pi/2$ rotation about the $y$ axis in
$su(2)_1$, leading to the identifications
\begin{eqnarray}
O_{-2,0}& \to &\;\;\; O_{0,\frac{1}{2}}=\;\;\; e^{\displaystyle{i\sqrt{2\pi} \widetilde \Phi}}\nonumber\\
O_{2,0} &\to& -O_{0,-\frac{1}{2}}=-e^{\displaystyle{-i\sqrt{2\pi} \widetilde \Phi}}\nonumber \\
O_{0,-\frac{1}{2}} &\to& -O_{2,0}=-e^{\displaystyle{i \sqrt{2\pi} \Phi}}\nonumber\\
O_{0,\frac{1}{2}} &\to& -O_{-2,0}=-e^{\displaystyle{-i\sqrt{2\pi} \Phi}}
\label{gtransf}
\end{eqnarray}
where the operators on the right hand side of Eq.~\eqref{gtransf} are
vertex operators written in terms of the new bosons $\Phi$ and
$\widetilde \Phi$. Notice that this is a duality
transformation.

However, there is an operator in this theory, namely $O_{\pm1,0}$, which
does not have $su(2)_1$ quantum numbers. As a result it will have a
quite different behavior along the phase boundary. In contrast, all the
other operators in this theory carry $su(2)_1$ quantum numbers. (As
remarked in Ref.~\onlinecite{ginsparg88} this is not truly an $su(2)_1$
theory, although it contains it, but rather connected with a
$\mathbb{Z}_2$ orbifold as well.) It is straightforward to check that
the operators $O_{\pm 1,0}$ do not have an OPE with the operators which
do transform under $su(2)_1$ (or, rather, that the OPE involve only
irrelevant operators.) Since the marginal operator which deforms the
$su(2)_1$ theory along the phase boundary does transform under
$su(2)_1$, the operator $O_{\pm 1,0}$ will not mix (in the sense of its
OPE) with the marginal operator either. We will see that this implies
that the dimension of this operator remains equal to $1/4$ along the
entire line, a result that is also well known (see for instance
Ref.~\onlinecite{Kohmoto81}.) In contrast, the operators of the
spin-$1/2$ representation, Eq.~\eqref{spin1/2}, do transform under
$su(2)_1$, a fact which is generated by their OPE's with the $su(2)_1$
generators\cite{yellow} and we will now see that their scaling
dimensions do change along the phase boundary line. In Section
\ref{sec:MC} we present evidence from Monte Carlo simulations in support
of both statements.

We can now use this approach to solve this problem exactly along the
phase boundary. Formally the Hamiltonian $\mathcal{H}_0$ of
Eq.~\eqref{Hrotated}, is equivalent to a spinless Luttinger model with
attractive backscattering interactions\cite{emery79}. As in the case of
the Luttinger model, the problem is solved by means of a Bogoliubov
transformation of the right and left moving bosons. This procedure
breaks the $su(2)_1$ symmetry explicitly. We introduce a new bose field
$\chi$ and its dual field $\widetilde \chi$. The left and right moving
components of these fields, $\chi_L$ and $\chi_R$, are linearly related
to the left and right moving fields $\Phi_L$ and $\Phi_R$ by
\begin{eqnarray}
\chi_L&=&\frac{1}{2}\left(\sqrt{\kappa}+\frac{1}{\sqrt{\kappa}}\right) \Phi_L+
	        \frac{1}{2}\left(\frac{1}{\sqrt{\kappa}}-\sqrt{\kappa}\right) \Phi_R
	       \nonumber\\
\chi_R&=&\frac{1}{2}\left(\frac{1}{\sqrt{\kappa}}-\sqrt{\kappa}\right) \Phi_L+
	        \frac{1}{2}\left(\sqrt{\kappa}+\frac{1}{\sqrt{\kappa}}\right) \Phi_R
	       \nonumber \\
&&
\label{chi-phi}
\end{eqnarray}
where $\kappa$ is given by
\begin{equation}
\kappa=\sqrt{\frac{1+2\pi (z+w)}{1-2\pi(z+w)}}
\label{kappa}
\end{equation}
The inverse transformation of Eq.~\eqref{chi-phi}, which relates
$\Phi_{L}$ and $\Phi_{R}$ to $\chi_{L}$ and $\chi_{R}$ has the same form
and it is obtained simply by replacing $\kappa$ by $1/\kappa$.

The Hamiltonian $\mathcal{H}_0$ in terms of the new fields becomes
\begin{equation}
\mathcal{H}_0=\frac{v}{2} \left[\left(\partial_x \widetilde \chi \right)^2+\left(\partial_x \chi\right)^2\right]
\label{H0}
\end{equation}
where the ``dimensionless velocity'' $v$ is
\begin{equation}
v=\sqrt{1-4\pi^2(z+w)^2}
\label{velocity}
\end{equation}
which can be absorbed in a suitable rescaling of the $x$ coordinate, $x
\to x \sqrt{v}$. Notice that the parameter $\kappa$ plays the same role
as the stiffness $K$ defined above, which governed the change of the
scaling dimensions at zero doping. Similarly, $\kappa$ governs the
change of the scaling dimensions along the $z=w$ phase boundary of the
systems at finite hole density. Here too, the relationship between this
stiffness $\kappa$ and the microscopic interactions is non-universal,
and the validity of Eqs.\ \eqref{kappa} and~\eqref{velocity} is
restricted to the weak coupling regime in which this continuum theory
holds. Notice that, since $z\geq 0$ and $w\geq 0$, we will always have
$\kappa \geq 1$. This fact will play an important role below.

We can now use these results to determine the scaling dimension of the
perturbation $\mathcal{H}_{\rm pert}$ along the phase boundary. It is
straightforward to write the perturbation $\mathcal{H}_{\rm pert}$ in
terms of the new field $\chi$ and its dual $\widetilde \chi$:
\begin{eqnarray}
\mathcal{H}_{\rm pert}
&=&-\frac{2(z-w)}{a^2} \sin(\sqrt{8\pi}\Phi_L) \sin(\sqrt{8\pi}\Phi_R)
\nonumber \\
&=&\frac{(z-w)}{a^2} \left[\cos\left( \sqrt{8\pi\kappa} \; \chi \right)-\cos\left(\sqrt{\frac{8\pi}{\kappa}}\; \widetilde \chi\right) 
\right]\nonumber\\
&&
\label{Hpert}
\end{eqnarray}

\subsection{Critical behavior along the phase boundary}
\label{sec:critical}

\subsubsection{The correlation length exponent}
\label{sec:correlation}

From Eq.~\eqref{Hpert} we find that the operator which perturbs the line
of fixed points along the phase boundary at finite density,
$\mathcal{H}_{\rm perp}$, involves two operators whose scaling
dimensions are $2\kappa>2$ and $\displaystyle{\frac{2}{\kappa}}<2$
respectively, since $\kappa >1$. Thus this operator becomes more
relevant along the phase boundary, away from the KT point, which in this
language has $\kappa=1$. In fact, if we neglect the effects of the
irrelevant operator (which is a safe thing to do only away from the KT
point since its only important effect is a finite renormalization of
$\kappa$) we see that the effective theory in the vicinity of the phase
boundary is a sine-Gordon theory for the dual field $\widetilde \chi$.
Since the scaling dimension of the relevant operator is $2/\kappa$, it
follows that, away from the KT point, the correlation length $\xi$
diverges as the phase boundary is approached as
\begin{equation}
\xi \sim |z-w|^{-\nu}, \qquad \nu=\frac{1}{2-\displaystyle{\frac{2}{\kappa}}}=\frac{\kappa}{2(\kappa-1)}
\label{xi-exponent}
\end{equation}
Thus, the correlation length exponent decreases (from infinity!) along
the phase boundary away from the KT point. It is apparent from the form
of the perturbation that away from the KT point there is simple scaling,
up to contributions of strictly irrelevant operators. On the other hand,
as $\kappa \to 1$ the relevant operator becomes marginally relevant and
the irrelevant operator becomes marginally irrelevant. Thus, as $\kappa
\to 1$ we should expect logarithmic corrections to scaling, and a
complex crossover near $\kappa=1$.

\subsubsection{The columnar order parameter and its susceptibility}
\label{sec:columnar}

On the other hand, we noted above that the dimension of the columnar
order parameter operator, $\frac{1}{2}(O_{1,0}+O_{-1,0})$ remains fixed
at the KT value of $\Delta_{1,0}=1/8$. Hence for this operator we find
$\eta_{1,0}=1/4$. On the other hand from scaling we know that the
susceptibility exponent obeys the scaling relation
$\gamma_{1,0}=(2-\eta) \nu$, where $\nu$ is given by
Eq.~\eqref{xi-exponent}. Hence
\begin{equation}
\gamma_{1,0}= \frac{7\kappa}{8(\kappa-1)}
\label{gamma-10}
\end{equation}
is the susceptibility exponent of the columnar order parameter, which
also increases along the phase boundary, even though
$\gamma_{1,0}/\nu=7/4$ along the whole phase boundary (provided the
transition remains continuous!.)

\subsubsection{The orientational order parameter and its susceptibility}
\label{sec:orientational}

We can use the operator identifications to look at the behavior of the
orientational order parameter which we saw above is the operator
$O_{2,0}$ of the original version of the theory. We also saw that this
operator is a component of the spin-$1/2$ representation of $su(2)_2$.
We also found how it transforms. In particular, we have
\begin{equation}
O_{\pm 2,0} \to \mp e^{\displaystyle{\mp i \sqrt{\displaystyle{\frac{2\pi}{\kappa}}}\;  \widetilde\chi}}
\label{O20-map}
\end{equation}
Along the phase boundary the scaling dimension of this operator is
\begin{equation}
\Delta_{2,0}=\frac{1}{2\kappa}< \frac{1}{2}
\label{dim20}
\end{equation}
which it is always relevant, and
\begin{equation}
\eta_{2,0}=\frac{1}{\kappa}
\label{eta20}
\end{equation}
Using once again the scaling relation $\gamma_{2,0}=(2-\eta_{2,0})\nu$, we find that the susceptibility exponent for the orientational order parameter is
\begin{equation}
\gamma_{2,0}=\frac{2\kappa-1}{2(\kappa-1)}
\label{gamma20}
\end{equation}
which also decreases along the phase boundary away from the KT point.

\subsection{Tricritical Point, First-Order Transition, and Phase Separation}
\label{first-order}

Let us now discuss how this critical line turns into a first-order
transition at a multicritical point. In Sections \ref{sec:MC} and
\ref{sec:1storder} we use Monte Carlo simulations to show that this is
indeed what happens. In Section \ref{sec:finite-density} we used mean-field
methods which indicated that the transition eventually should become
first order. For this to work we should be able to predict the existence
of a tricritical point along the phase boundary at which the transition
becomes first order. It turns out that this is the case and that the
first-order transition is triggered by an effective attractive
interaction between holes on the same sublattice, leading to phase
separation.

To see how this happens we need to discuss the effects of irrelevant
operators along the phase boundary.  As we stated above, their most
important effect is a finite and non-universal renormalization of
$\kappa$ away from the value given in Eq.~\eqref{kappa}. We have also
focused on the role of the operators $O_{4,0}$ and $O_{0,1}$ as they are
both marginal at the KT transition. However, we also saw that one
combination of these two operators remains marginal along the phase
boundary and its coupling constant determines the value of $\kappa$,
through Eq.~\eqref{kappa}. On the other hand, the other combination is
the sum of a relevant operator, $\cos(\sqrt{(8\pi/\kappa)} \widetilde
\chi)$ with scaling dimension $2/\kappa$, and of an irrelevant operator,
$\cos(\sqrt{8\pi \kappa} \chi)$ with scaling dimension $2\kappa$. The
dimension of the irrelevant operator increases along the phase boundary
(thus becoming more irrelevant) while the dimension of the relevant
operator decreases as $\kappa$ increases (thus becoming more relevant.)

One possible scenario for a first-order transition is found by noting
that as $\kappa \to \infty$, the dimension of the relevant operator
vanishes, and the ``thermal eigenvalue'' $\gamma_{0,1}/\nu \to 2$. Thus
at this point, naturally provided this limit is accessible, the line of
fixed points reaches a discontinuity fixed point\cite{nienhuis76} and
the transition becomes first order. However, at this point the theory
becomes pathological (as $\kappa \to \infty$, $v \to 0$) and one may
suspect that other physical effects, contained in irrelevant operators,
may intervene before this happens.

We have so far neglected other operators which are even more irrelevant
at the KT transition. For example, the operators $O_{8,0}$ and
$O_{0,2}$, have dimension $8$ at the KT point. Recall that the operator
$O_{0,2}$ represents pairs of holes on the {\em same sublattice}. Both
of these operators are present in any lattice problem (such as the
interacting dimer model) and play no significant role at the KT
transition (beyond a non-universal but otherwise trivial shift of the
critical coupling) and for this reason they were (correctly) neglected.
However, along the phase boundary the scaling dimensions of these
operators change. Using the OPE, it is easy to see that along the phase
boundary both operators contain the operators (among others which are
less important) $O_{0,2} \sim \cos(2\sqrt{(8\pi/\kappa)} \widetilde
\chi)$ with scaling dimension $8/\kappa$, and $O_{8,0} \sim
\cos(2\sqrt{8\pi \kappa} \chi)$ with scaling dimension $8\kappa$. Even
though they are not explicitly present in our starting theory, these
operators will be generated under renormalization and close enough to
the KT point, $\kappa \gtrsim 1$, they both are and remain irrelevant.

However, although $\kappa$ also changes in a non-universal manner, the
dependence of the dimensions with $\kappa$ does not, as it follows from
the structure of the theory. Thus, provided the dependence between
$\kappa$ and the microscopic couplings allow it, it may be possible to
reach a point along the phase boundary at which $\kappa=4$. This will
happen at a critical value of the coupling constant $u$ and a critical
value of the hole density $\rho$ (or, equivalently at a critical value
of the hole fugacity $z$ (cf.\ Fig.~\ref{fig:sketch}) ).

At this critical value of $\kappa$, the scaling dimension of the operators $O_{0,2}$ becomes equal to $2$, and together with the strictly marginal operator $O_{4,0}$, there are now two marginal operators at this point. Thus the system is at a {\em tricritical} point at this value of the parameters.\cite{Bruce75} 
Past this point, $O_{0,2}$ becomes marginally relevant along the phase
boundary. In this regime, the effective field theory at the phase
boundary is a sine-Gordon theory for the field $\tilde \chi$ with the
marginally relevant operator $O_{0,2}$. Since the sine-Gordon theory in
this regime is massive, it has a finite correlation length and since
$O_{0,2}$ is marginally relevant the correlation length along the phase
boundary, which has now become a coexistence curve, has an essential
singularity as a function of the distance to the tricritical point, {\it
i.e.\/} a KT-like transition. Thus, the transition becomes {\em first
order} along the phase boundary past the tricritical point with a
correlation length that scales like $\xi\sim e^{{\rm const}./\sqrt{s}}$,
where $s$ is the distance to the tricritical point measured along the
coexistence curve. In contrast, the correlation length across the phase
boundary (below the tricritical point) exhibits conventional power-law
scaling. 
Closely related scenarios for the existence of such tricritical points have been suggested in other
systems, such as the extended Hubbard model in
one-dimension\cite{fradkin-Hirsch83}, the two-dimensional classical
Ashkin-Teller model\cite{Grest81,Fradkin84}, and the dilute 4-state Potts model\cite{cardy80}, which is a statistical system with very similar phase diagram.

What happens as the tricritical point is reached, can be understood more
physically by noting that at that point
the operator $O_{0,2}$, which measures the probability amplitude for a pair of holes
(on the same sublattice), becomes relevant. The relevance of $O_{0,2}$
indicates that holes on the same sublattice now have a strong effective
attractive interaction, have a strong tendency to pairing and consequently phase
separate. The effective field theory description given above corresponds
to the grand-canonical picture, since the coupling constants are simple
functions of the hole fugacity. On the other hand, in the canonical
description, {\it i.e.\/} at fixed hole density $\rho$, the coexistence
curve opens up into a two-phase region: there is phase separation
between hole-poor regions with local columnar dimer order and hole-rich
regions. The jump in the hole and dimer densities (as well as in the
order parameters) across the first-order transition is governed by the
correlation length at the coexistence curve. Thus, close to the
tricritical point the jump in the densities ({\it i.e.\/} the width in
density of the two-phase region) has the scaling form $\Delta \rho \sim
\xi^{-2}$ and therefore vanishes with an essential singularity as the tricritical
point is approached. Similar scaling behavior applies to the
discontinuity of the columnar and orientational order parameters across
the two-phase region.

In the subsequent sections we will give further evidence for the nature
of the phase transitions in this system, including the first-order
transition, using Monte Carlo simulations in the canonical and grand-canonical ensemble.

\section{Monte Carlo Simulations}
\label{sec:MC}
 
We now employ Monte Carlo simulations to map out the phase diagram of
the doped quantum dimer models at their generalized RK points.
This approach is complementary to the analytic approach of Sections
\ref{sec:mean-field}, \ref{sec:zero-density}
and~\ref{sec:finite-density}, and of Appendix
\ref{app:mean-field-details}. We first introduce (subsection
\ref{sec:methods}) a canonical Monte Carlo algorithm for interacting
dimers.  In subsection \ref{sec:zero-doping} we apply this method to the
case of the interacting fully packed classical dimer model on the square
lattice, a system that has been studied recently in some
detail,\cite{alet05,alet06,Castelnovo06,Poilblanc06} and study its phase
transition. In subsections \ref{sec:low-doping} and~\ref{sec:1storder}
we consider the case of the doped dimer model at low doping and map out
the critical line, verifying the theoretical scenario discussed in
Section~\ref{sec:finite-density}. In subsection~\ref{sec:1storder} we
combine the cluster algorithm with conventional grand-canonical moves
that permit us to determine the first-order transition line as a
function of dimer fugacity~$z_d$ and interaction strength~$u$. We also
estimate the location of the multicritical point discussed in
Appendix~\ref{app:mean-field-details} and Section
\ref{sec:finite-density}, using data from both canonical and
grand-canonical simulations.
  
\subsection{Algorithm for classical interacting dimers}
\label{sec:methods}

At high dimer coverage (low doping), conventional Monte Carlo algorithms
become very inefficient. On the other hand, in
Ref.~\onlinecite{krauth03}, it was demonstrated that a geometric cluster
algorithm (GCA) can work efficiently for dimers that only have a
repulsive hard-core interaction. We briefly summarize this algorithm
here.  The overlap of two hard-core dimer configurations generates a
transition graph.  This graph consists of disjoint subgraphs of dimers
alternating between the two configurations. In the presence of holes,
there are two possible types of graphs: an \emph{open graph} which
always terminates on a hole or a \emph{closed loop}.  Any Monte Carlo
move corresponds to a transition graph of the initial and final
configurations.
In the geometric cluster algorithm, the two subgraphs are related by a
global lattice symmetry. The algorithm obtains long transition graphs
with minimal overhead: moves are never rejected, and each dimer
encountered during the construction of the graph participates in the
move.  The construction proceeds as follows.\cite{krauth03} First, a
``seed'' dimer and a symmetry axis are chosen at random. The seed dimer
is reflected with respect to the symmetry axis, and if it overlaps with
other dimers these are reflected as well. This proceeds in an iterative
fashion until there are no more dimer overlaps or, equivalently, when an
open or closed graph has been formed.  On the square lattice, the
algorithm is ergodic if we allow both diagonal and horizontal-vertical
axes passing through sites of the lattice. The first choice allows to
change the numbers of horizontal and vertical dimers, whereas the second one
permits to move through the different winding number sectors.
Transition graphs generated by the algorithm are symmetric with respect
to the symmetry axis, and cross it at most twice.

We now extend this approach to dimers with additional interactions by
exploiting the \emph{generalized} geometric cluster algorithm proposed by Liu
and Luijten.\cite{liu04,liu05a} Now, in a single cluster move, multiple
transition graphs and/or open graphs are formed simultaneously while
retaining the rejection-free character of the algorithm. This is
achieved by also reflecting dimers that do \emph{not} overlap,
with a probability that depends on the dimer-dimer coupling.  When a
dimer~$i$, located at $\vec r^{\rm old}_i$, is reflected to a new
position $\vec r^{\rm new}_i$, there are two classes of dimers that
interact with dimer~$i$: a) dimers which interact with it
\emph{before} it is reflected and b) dimers which interact with~$i$
\emph{after} it is reflected. Dimers~$j$, located at positions $\vec
r_j$, that belong to any of the two classes are included in the cluster
(\ie will be reflected with respect to the symmetry axis) with a
probability
\begin{equation} 
 p_{ij}=\max \left[1-e^{-\frac{\delta\mathcal{U}_{ij}}{k_BT}},0\right]
 \;,
\end{equation}
where $\delta\mathcal{U}_{ij} =V(|\vec{r}^{\rm new}_i - \vec{r}_j|)-V(|
\vec{r}^{\rm old}_i -\vec{r}_j|)$ and $V(r)$ represents the interaction
between two dimers at a separation~$r$.  Thus the cluster addition
probability for dimer~$j$ depends \emph{solely} on the energy difference
corresponding to a change in relative position of $i$ and~$j$. In the
limit of a pure hard-core repulsion, this generalized geometric cluster
algorithm reduces to the original GCA.

The GGCA applies only to simulations in the canonical ensemble. To
perform Monte Carlo simulations in the grand-canonical ensemble, we
alternate the cluster moves with conventional grand-canonical Metropolis
moves, consisting of insertion and deletion attempts of single dimers.

\subsection{Zero doping:  Kosterlitz-Thouless transition to a columnar valence-bond crystal}
\label{sec:zero-doping}

\begin{figure}[hbt]
\includegraphics[width=0.45\textwidth]{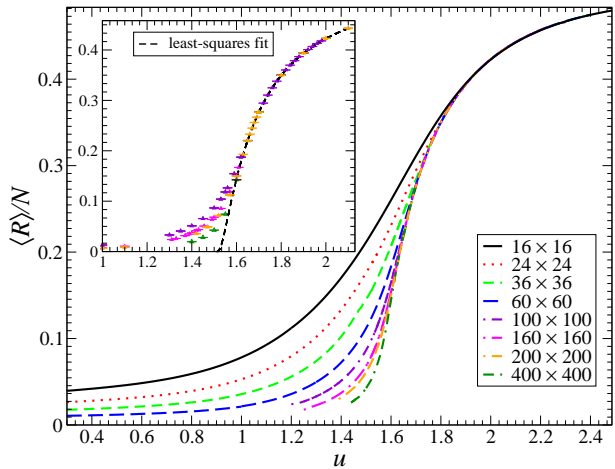}
\caption{(color online) The orientational order parameter $R$ in the
undoped case: it vanishes for $u<u_c$ and has an essential singularity
at $u_c$. The curves represent Monte Carlo data for different lattice
sizes, interpolated via multiple histogram reweighting.  Inset: data
collapse, with a least-squares fit to the exact scaling function for the
staggered polarization operator of the six-vertex
model~\eqref{baxterscaling} related to $R$ by a universality mapping as
discussed in the text.}
\label{fig:orderpar}
\end{figure}

According to the theoretical study of Section~\ref{sec:zero-density} and
also from the results of Refs.~\onlinecite{alet05,alet06,Castelnovo06},
we expect to find a Kosterlitz-Thouless transition at zero doping, as a
function of the dimer interaction.  To detect and locate this
transition, we exploit the fact that there is an ordered phase in the
large-$u$ region and define columnar and orientational order parameters,
\begin{eqnarray}
  {C}({\bf r}) &\equiv & \sum_{i=x,y} \left[ n_i({\bf r})-n_i({\bf
    r}+{\bf e}_i)\right] \;, \\
  {R}({\bf r}) &\equiv & n_x({\bf r})n_x({\bf r}+ {\bf e}_y)-n_y({\bf
  r})n_y({\bf r}+{\bf e}_x) \;,
  \label{orderpar}
\end{eqnarray}
where $n_i ({\bf r})$ denotes the dimer density at ${\bf r}$. $\langle
C({\bf r}) \rangle$ is non-vanishing only in a columnar-ordered phase,
providing a signature of translational symmetry breaking (with a
four-fold degeneracy), whereas $\langle R ({\bf r})\rangle $ measures
the breaking of invariance under $\pi/2$ rotations. In terms of the most
relevant operators of the effective theory of Section
\ref{sec:zero-density}, using OPE, we make the following
identifications:
\begin{eqnarray}
C &\sim&\frac{1}{2}(O_{1,0} + O_{-1,0}) \;, \\
R&\sim&\frac{1}{2}(O_{2,0}+O_{-2,0}) \;.
\end{eqnarray}
Having a proper correspondence between the effective theory described in
Section \ref{sec:zero-density} and the microscopic order parameters
\eqref{orderpar}, we may verify our predictions.  Since the conventional
fourth-moment ratio (directly related to the Binder
cumulant\cite{Binder81}) of the order parameter generally does not show
a well-defined crossing at a KT transition, we instead use a scaling
function of the form of the spontaneous staggered polarization of the
six-vertex model,\cite{baxter73} which maps on the same vertex operator
as $R$ in the Coulomb-gas representation, and is in the same
universality class. Keeping only the most relevant terms, this function
has the form,
\begin{equation}
  \langle R(u)\rangle =
  \left(a\frac{1}{\sqrt{u-u_c}} +
    \cdots\right)e^{\left[-\frac{c}{\sqrt{u-u_c}} +
    d\sqrt{u-u_c} + \cdots \right]} \;.
\label{baxterscaling}
\end{equation}
From a careful nonlinear least-squares fit to the numerical data outside
the finite-size regime (cf.\ Fig.~\ref{fig:orderpar}) we obtain
$u_c=1.508 \pm 0.003$.

\begin{figure}[hbt]
\includegraphics[width=0.45\textwidth]{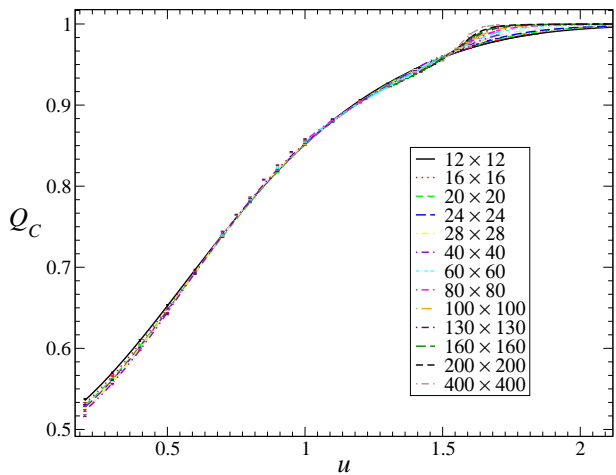}
\caption{(color online) Fourth-order amplitude ratio of the columnar
order parameter $C$ in the undoped case. The curves for all system sizes
essentially coincide for the entire critical phase ($u<u_C$), as
expected for a KT transition.}
\label{fig:CSB_cum_0}
\end{figure}

\begin{figure}[hbt]
\includegraphics[width=0.45\textwidth]{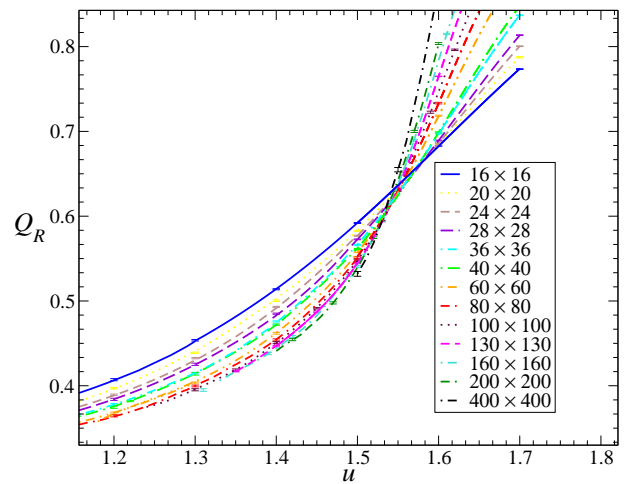}
\caption{(color online) Fourth-order amplitude ratio of the
orientational order parameter $R$ in the undoped case.  In contrast to
the amplitude ratio of the columnar order parameter $C$
(Fig.~\ref{fig:CSB_cum_0}), this quantity exhibits a strong finite-size
dependence, leading to an effective crossing point that can be used to
locate the transition point.}
\label{fig:PSB_cum_0}
\end{figure}

For completeness, we also investigate the behavior of the fourth-order
amplitude ratios $Q_{M} = \langle{M}^2\rangle^2/\langle {M}\rangle^4$
with $M = C$, $R$.  The behavior of $Q_C$, shown in
Fig.~\ref{fig:CSB_cum_0}, is similar to what is expected for the $XY$
model, namely a collapse of all curves in the critical low-$u$ phase and
no well-defined crossing of curves for different system sizes.  In
contrast, $Q_R$ (Fig.~\ref{fig:PSB_cum_0}) is found to exhibit such
strong finite-size effects in the critical phase that its behavior
almost resembles that of a regular continuous phase transition. This
anomalous behavior explains why the crossing point of the curves for
different system sizes could be exploited to obtain an accurate estimate
of the critical coupling\cite{alet05,alet06}. In the figures presented
in this section, the multiple histogram reweighting
method\cite{ferrenberg89} has been used to interpolate all data obtained
at different values for the coupling parameter. This allows us to
accurately locate crossing points and extrema in the curves.

Alet and coworkers\cite{alet05,alet06} and Poilblanc and coworkers\cite{Poilblanc06} 
used transfer-matrix calculations and Monte Carlo
simulations to study the critical behavior of the undoped system for
$u>0$ (attractive dimer interactions), whereas Castelnovo and
coworkers\cite{Castelnovo06} used transfer-matrix methods to study
primarily the $u<0$ (``repulsive'') regime. In addition, in
Ref.~\onlinecite{alet06} the doped interacting dimer model was also briefly
studied for low doping by means of numerical transfer-matrix
techniques. All these results are consistent and complementary to those
presented in the following subsection.

\subsection{Low doping: Line of fixed points}
\label{sec:low-doping}

\begin{figure}[t]
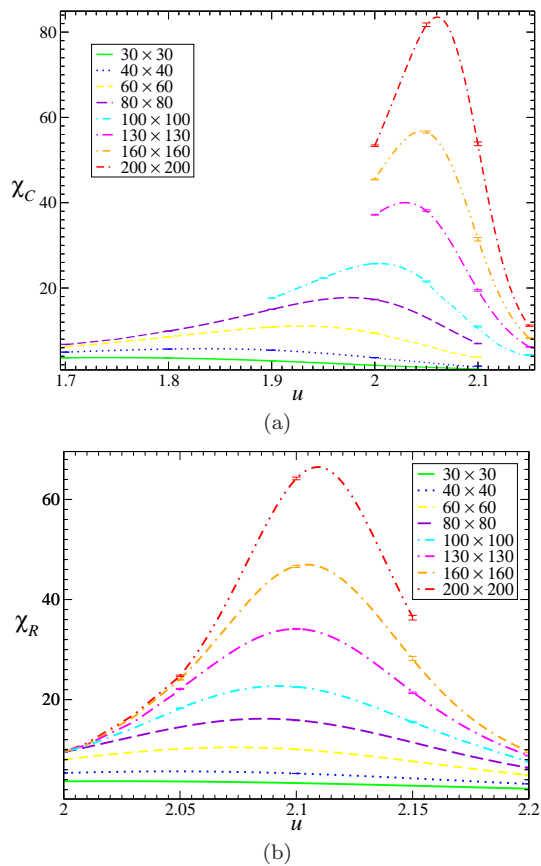

\subfigure[]{\includegraphics[width=0.39\textwidth]{csb_susc_scal_0.94.eps}
\label{susc1}}
\subfigure[]{\includegraphics[width=0.39\textwidth]{psb_susc_scal_0.94.eps}\label{susc2}}
\caption{(color online) Classical susceptibilities of (a) the columnar
order parameter $C$ and (b) the orientational order parameter $R$, for
$\rho_{h}=0.06$. The magnitude and position of the maxima for different
lattice sizes $L$ can be used to extract the critical exponents $\gamma$
and $\nu$.}
\label{susc}
\end{figure}

\begin{figure}[hbt]
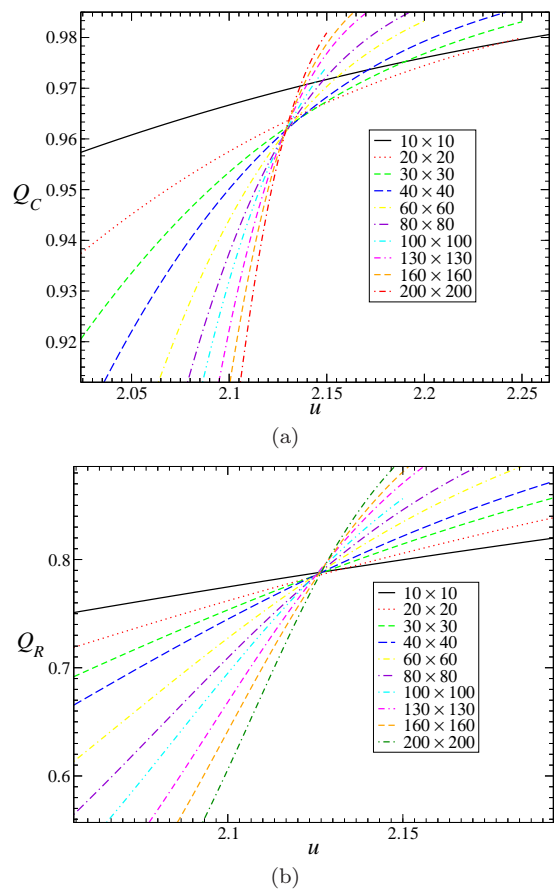

\subfigure[]{\includegraphics[width=0.4\textwidth]{csb_cumul_0.94.eps}
\label{cumul1}}
\subfigure[]{\includegraphics[width=0.4\textwidth]{psb_cumul_0.94.eps}}
\caption{(color online) Fourth-order amplitude ratios of (a) the
columnar order parameter $C$ and (b) the orientational order parameter
$R$, for $\rho_{h}=0.06$. The clear crossings of the curves for
different system sizes indicate a regular continuous phase transition.}
\label{cumul}
\end{figure}

We now proceed to the low-doping regime. We perform simulations in the
canonical ensemble, for couplings near the critical region and for hole
densities $\rho_{h}=0.004, 0.01, 0.02, 0.04, 0.06$. Figure~\ref{susc}
shows, for $\rho_{h}=0.06$, the susceptibilities of the columnar and
orientational order parameters, $C$ and $R$, followed by the
corresponding fourth-order amplitude ratios in Fig.~\ref{cumul}.

According to the predictions of Section~\ref{sec:finite-density}, we
expect that the columnar order parameter $C$, which maps onto the
$O_{1,0}$ effective operator in the scaling limit, will retain its
scaling dimension $1/8$ along the critical line that emerges from the KT
transition point for low hole doping and which constitutes the phase
boundary between the dimer-hole liquid and the columnar solid phases.
This constant value of the scaling dimension of $C$
is the most salient signature of this critical line: The scaling
dimensions of all the other operators change continuously along the
phase boundary.

\begin{figure}[hbt]
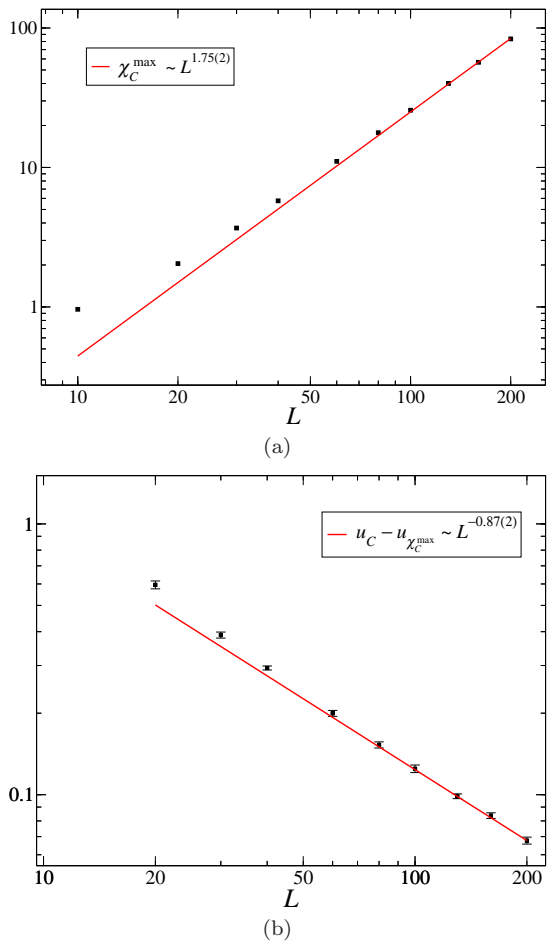

\subfigure[]{\includegraphics[width=0.4\textwidth]{csb_susc_max_scal_0.94.eps}}
\subfigure[]{\includegraphics[width=0.4\textwidth]{csb_susc_pos_scal_0.94_2.eps}\label{susc1pos}}
\caption{(color online) Finite-size scaling analysis of the
susceptibility of the columnar order parameter $C$ for $\rho_h=0.06$:
(a) Scaling of the height of the maximum as a function of system
size~$L$. (b) Scaling of the position of the maximum. The critical value
$u_c$ is obtained from the crossing of the fourth-order cumulants (cf.\
Fig.~\ref{cumul}). The same power-law behavior of the susceptibility
maximum is found for all hole densities $\rho_{h} \leq 0.06$.}
\label{C-susc}
\end{figure}

\begin{figure}[hbt]
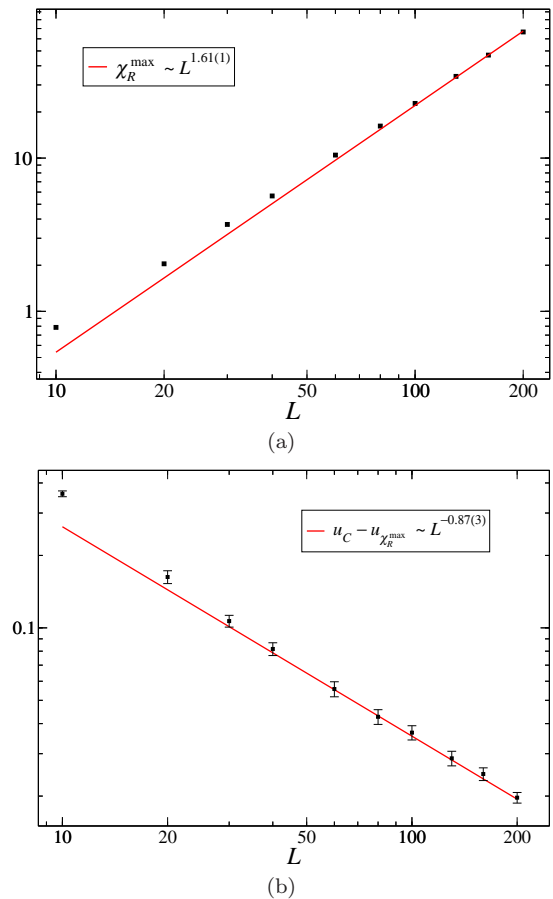

\subfigure[]{\includegraphics[width=0.4\textwidth]{psb_susc_max_scal_0.94.eps}}
\subfigure[]{\includegraphics[width=0.4\textwidth]{psb_susc_pos_scal_0.94_2.eps}\label{susc2pos}}
\caption{(color online) Finite-size scaling analysis of the
susceptibility of the orientational order parameter $R$ for
$\rho_h=0.06$: (a) Scaling of the height of the maximum as a function of
system size~$L$. (b) Scaling of the position of the maximum. The
critical value $u_c$ is obtained from the crossing of the fourth-order
cumulants (cf.\ Fig.~\ref{cumul}). Unlike the results for the columnar
order parameter (Fig.~\ref{C-susc}) the power-law behavior of the
susceptibility maximum is is now dependent on the hole density,
consistent with an anomalous scaling exponent that varies along the
phase boundary.}
\label{R-susc}
\end{figure}

To test this prediction, we extract the anomalous dimensions $\eta_C$
and $\eta_R$ (which are equal to twice their scaling dimension) by means
of finite-size scaling. The maximum of the susceptibility scales as
$\chi^{\rm max}\sim L^{2-\eta}+\cdots$. Subleading scaling contributions
are omitted in the fitting expression since the results for sufficiently
large lattice sizes satisfy simple scaling [cf.\ Figs.\ \ref{C-susc}
and~\ref{R-susc}].

The correlation-length exponent~$\nu$ can be extracted from the slope of
the fourth-order amplitude ratios of both order parameters at the
critical point. Here, instead, we obtain it from the scaling behavior of
the location of susceptibility maximum, which scales as
$u_{\chi^{\rm max}}=u_c + {\rm const.}\; L^{-1/\nu}+\cdots$ (cf.\ Figs.\
\ref{susc1pos} and~\ref{susc2pos}). The critical coupling $u_c$, in
turn, is obtained from the crossing points of the fourth-order amplitude
ratio (Fig.~\ref{cumul}).

\begin{figure}[hbt]
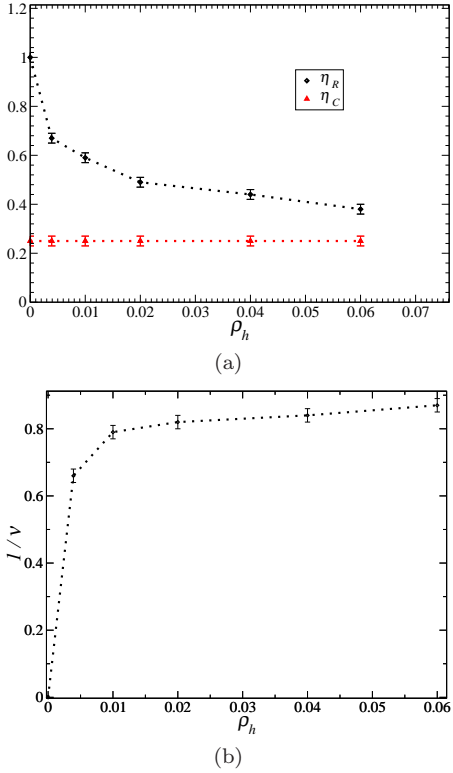

\subfigure[]{\includegraphics[width=0.33\textwidth]{gamma_nu.vs.rho2.eps}\label{exponents1}}
\subfigure[]{\includegraphics[width=0.33\textwidth]{nu_exp_low_doping2.eps}\label{exponents2}}
\caption{(a) Evolution of the anomalous dimensions $\eta_C$ (lower
curve) and $\eta_R$ (upper curve) for the columnar and orientational
order parameters, respectively, along the critical line for hole densities
$0.004 \leq \rho_{h} \leq 0.06$.  The transition will become first-order when the
two curves cross.  (b) The inverse correlation-length exponent $1/\nu$ as a
function of doping along the critical line for hole densities
$0.004 \leq \rho_{h} \leq 0.06$. Up to small systematic deviations discussed in
the text, the observed trend agrees with the scaling
predictions.}
\end{figure}

By repeating this procedure for different hole densities, we find
$\eta_C$ and $\eta_R$, as well as $\nu$ as a function of $\rho_h$. We
note that for the \emph{undoped} case the (logarithmic) finite-size
corrections are so strong that the anomalous exponents are very
difficult to determine. By including subleading corrections to the
susceptibility expressions at the KT transition, we find strong
indications that $\eta_{C}=1/4$ and $\eta_{R}=1$, satisfying the
established theoretical description of this KT transition.
The density dependence of the exponents is shown in
Fig.~\ref{exponents1}. Clearly, they behave very differently: For the
columnar parameter $C$, its anomalous dimension remains unchanged and equal
to $1/4$, while for $R$ it decreases monotonically from $1$ to
$1/4$ where the transition is expected to become first-order,
according to the scenario presented in Section \ref{sec:finite-density}.
The results from our Monte Carlo simulations are thus
consistent with the predictions we made in our theoretical analysis.

The evolution of the correlation length exponent along the phase
boundary is shown in Fig.~\ref{exponents2}. This exponent behaves
\emph{qualitatively} as predicted for finite doping, \ie it exhibits a
monotonically decreasing behavior along the line of fixed points.  A
direct \emph{quantitative} comparison to the field-theoretical
prediction is not possible, since the simulations are performed in the
canonical ensemble. Even though the dimer density could be mapped to a
\emph{dimer} fugacity, the field theory assumes a fixed \emph{hole}
fugacity.  In addition, the evolution of the exponent in
Fig.~\ref{exponents2} is slower than predicted because in the
simulations we do not approach the phase boundary perpendicularly in the
simulations, which leads to an \emph{effective} exponent $\nu$ that is
smaller than the one computed from the scaling dimension of the relevant
operator in the field theory.

\begin{figure}[hbt]
\includegraphics[width=0.34\textwidth]{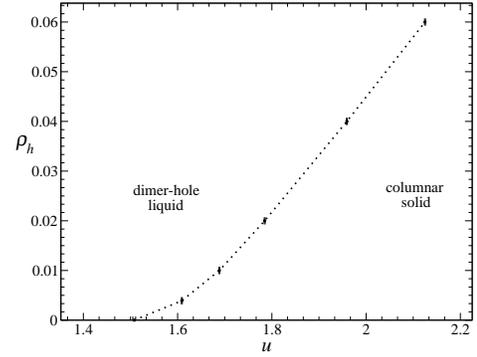}
\caption{Phase diagram at low doping, with the critical line of fixed
points separating the dimer-hole liquid phase
[$u<u_c(x)$] from the columnar solid [$u>u_c(x)$].}
\label{rhojc}
\end{figure}

In Fig.~\ref{rhojc} we summarize our results for the location of the phase
boundary between the dimer-hole liquid phase and the columnar solid
phase for hole densities $\rho_h \leq 0.06$.  The behavior of the
critical line for $\rho_h \to 0$ is consistent with its expected scaling
behavior $\rho_h \propto \xi^{-2} (\rho_h=0)$ which is based on simple
dimensional analysis, with the only length scale of the problem being
the correlation length of the undoped problem.

\subsection{High doping: First Order Transition and Phase Separation}
\label{sec:1storder}

Beyond the multicritical point predicted in Sections
\ref{sec:mean-field} and~\ref{sec:finite-density} (see also Appendix
\ref{app:mean-field-details}), we expect a first-order transition line
in the $z_d$--$u$ phase diagram, where $z_d$ denotes the dimer fugacity.
This line separates the crystalline from the liquid phase. According to
our scenario, an entropic attraction between holes on the same
sublattice becomes marginal and leads naturally to phase separation
between a hole-rich liquid phase and a hole-poor columnar crystalline
phase. Since the multicritical point is characterized by a marginally
relevant operator, complicated crossover will be observed in the
first-order transition region in the vicinity of this
point.\cite{cardy80} In addition, close to the multicritical point, the
first-order transition will be very weak, with a discontinuity that
vanishes with an essential singularity as a function of the distance to
the multicritical point along the phase coexistence curve. At this
transition, all observables, such as the latent heat, should vanish in a
similar way, making the numerical study of the transition close to the
multicritical point particularly difficult.

\begin{figure}[hbt]
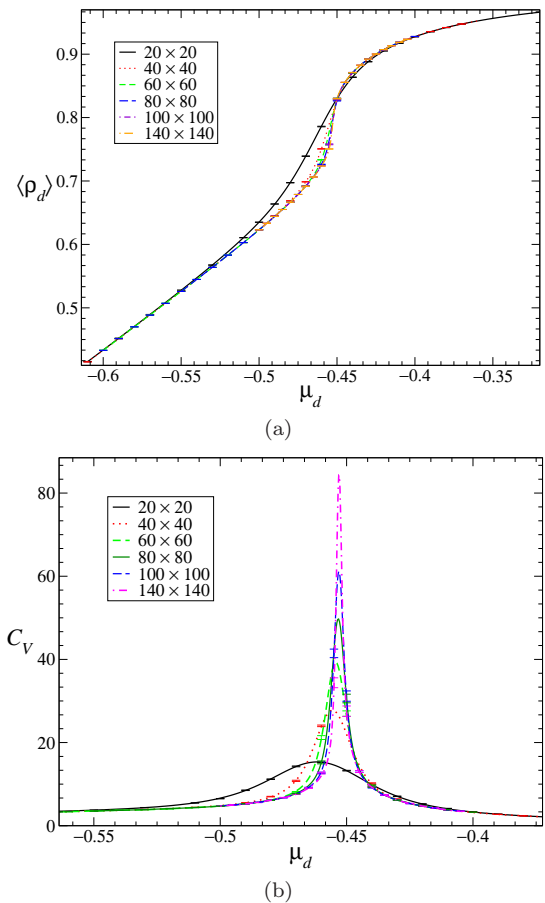

\subfigure[]{\includegraphics[width=0.39\textwidth]{dimerd_vs_cp_J3.50.eps}\label{dd}}
\subfigure[]{\includegraphics[width=0.4\textwidth]{heat_capac_vs_cp_J3.50.eps} \label{heatcap}}
\caption{(color online) (a) The $\rho_d$--$\mu_d$ equation of state for
coupling $u=3.5$ and different lattice sizes. As the system size
increases, the jump in the dimer density at the first-order phase
transition gradually becomes more pronounced, but remains relatively
weak. (b) The heat capacity of the classical interacting dimer model for
the same coupling strength. We find that the peak scales as $C_{\rm max}
\sim C_0 + L^2$, providing strong evidence for a first-order transition
at $u=3.5$.}
\label{ddheat}
\end{figure}

\begin{figure}[hbt]
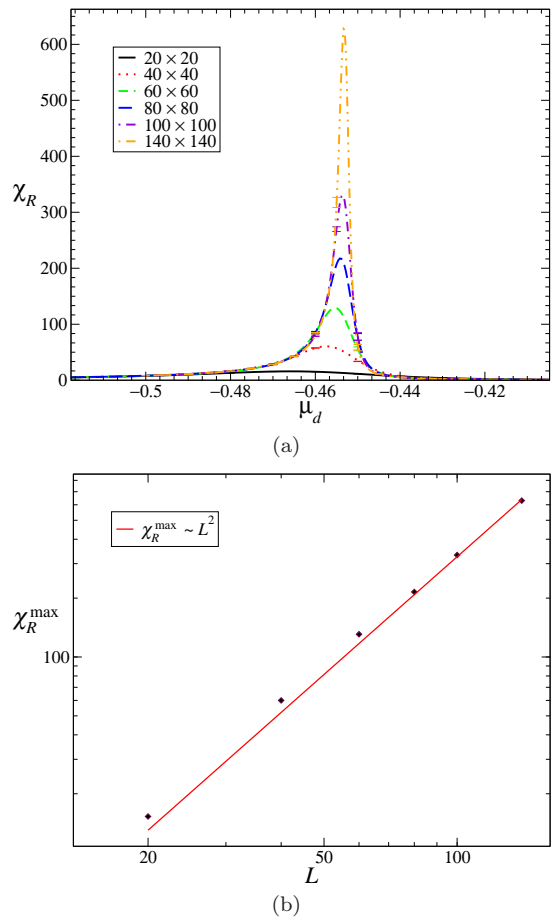

\subfigure[]{\includegraphics[width=0.4\textwidth]{psb_susc_vs_cp_J3.50.eps} \label{susc1fo1}}
\subfigure[]{\includegraphics[width=0.4\textwidth]{psb_suscmax_vs_L_J3.50.eps} \label{susc1fo2}}
\caption{(color online) (a) Susceptibility of the orientational order
parameter~$R$ for coupling $u=3.5$ and different lattice sizes.  (b)
Asymptotically, the maxima of $\chi_R$ scale with the lattice size,
confirming the first-order nature of the phase transition.}
\label{susc1fo}
\end{figure}
 
\begin{figure}[hbt]
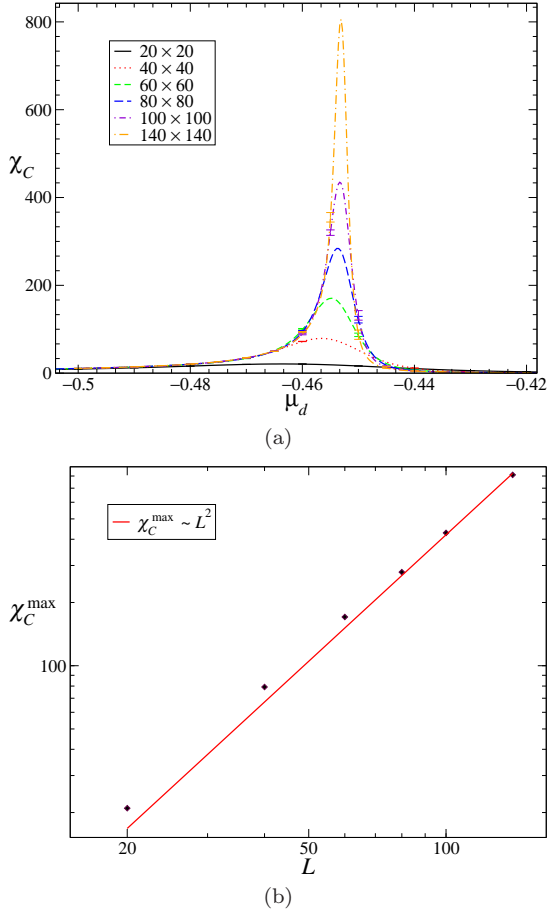

\subfigure[]{\includegraphics[width=0.4\textwidth]{csb_susc_vs_cp_J3.50.eps}\label{susc2fo1}}
\subfigure[]{\includegraphics[width=0.4\textwidth]{csb_suscmax_vs_L_J3.50.eps}\label{susc2fo2}}
\caption{(color online) (a) Susceptibility of the columnar order parameter~$C$ for
coupling $u=3.5$.  (b) The maxima of $\chi_C$ as a function of the linear lattice dimension. As
for $\chi_R$ (Fig.~\ref{susc1fo2}), the maxima scale with lattice size,
supporting the occurrence of a $\delta$-function singularity in $\chi_C$ in the
thermodynamic limit, as expected for a first-order phase transition.}
\label{susc2fo}
\end{figure}

To confirm the existence of the discontinuous transition we perform
grand-canonical Monte Carlo simulations with single-dimer insertions and
deletions (alternated with canonical geometrical cluster moves to
accelerate the relaxation of the configurations), for couplings $u=
3.0$, $3.5$, $4.0$, $5.0$, $6.0$, as a function of the dimer fugacity
$z_d$.  Figure~\ref{dd} shows the dimer density as a function of dimer
chemical potential~$\mu_d$. Although with increasing system size a jump
in the dimer density develops, it does not become very pronounced.
However, consideration of the heat capacity $C_V$ [Fig.~\ref{heatcap}]
confirms the presence of a single phase transition at fixed coupling
constant~$u$, as $C_V$ exhibits a peak at a chemical potential that
matches the location of the jump in $\rho_d$. We emphasize that this
classical heat capacity is \emph{not} the heat capacity of the
$(2+1)$-dimensional QDM\@. Indeed, $C_V$ does not have any physical
meaning in terms of the ground-state wave function that we are
considering, because the ground-state energy cannot be changed through
variation of the parameters $u$ or $\mu_d$. Another confirmation of the
phase transition is obtained from the susceptibilities of the
orientational [Fig.~\ref{susc1fo1}] and columnar [Fig.~\ref{susc2fo1}]
order parameters, $\chi_R$ and $\chi_C$, respectively. Both quantities
exhibit a peak at a chemical potential that approaches, with increasing
system size, the location of the peak observed in $C_V$.

\begin{figure}[hbt]
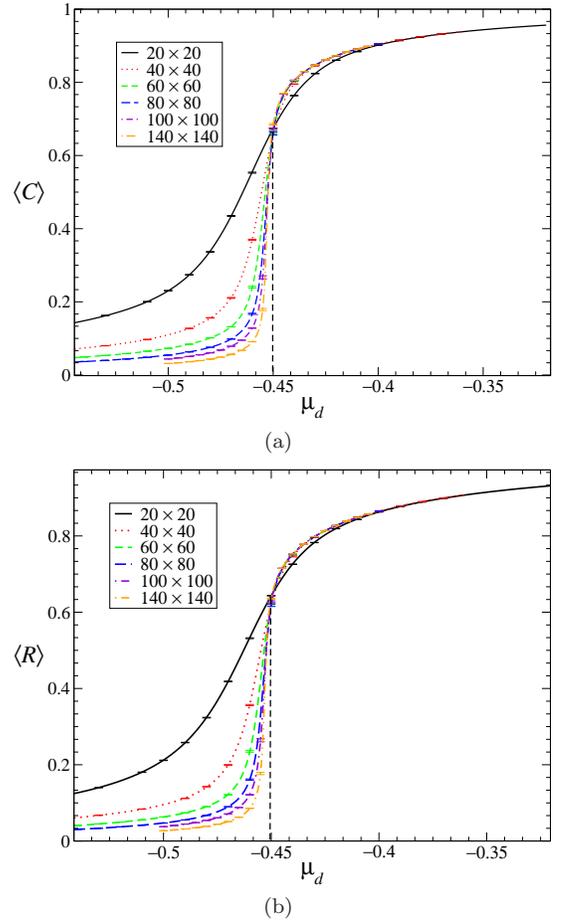

\subfigure[]{ \includegraphics[width=0.4\textwidth]{csb_op_vs_cp_J3.50.eps}\label{op1}}
\subfigure[]{ \includegraphics[width=0.4\textwidth]{psb_op_vs_cp_J3.50.eps}\label{op2}}
\caption{(color online) (a) The columnar order parameter $C$ as a
function of the dimer chemical potential $\mu_d$ for coupling $u=3.5$.
The rapid increase near the transition becomes more pronounced as the
thermodynamic limit is approached. The dashed line shows the approximate
position of the transition point.  (b) The orientational order parameter
$R$ as a function of the chemical potential $\mu_d$ for coupling
$u=3.5$. It also shows a rapid increase similar to the behavior of $C$,
at the same chemical potential.}
\label{op}
\end{figure}

To confirm the nature of the phase transition, we consider the scaling
of the peaks in $C_V$, $\chi_R$, and $\chi_C$. For a first-order
transition, these quantities should exhibit a $\delta$-function
singularity in the thermodynamic limit or equivalently, for finite
systems their peaks must scale with the lattice size in a finite system.
We find that the heat-capacity peak, apart from a constant background,
indeed scales with the lattice size~$L^2$ for the range of system sizes
(up to $L=140$) that we considered. This is supported by the behavior of
the system-size dependent maxima in $\chi_R$ and $\chi_C$, see Figs.\
\ref{susc1fo} and~\ref{susc2fo}, which both scale as $L^2$, indicating
that $\eta = 0$. In addition, all local order parameters must develop a
discontinuity at the transition point as the system size increases.
Whereas the jump in the dimer density [Fig.~\ref{dd}] is not very sharp,
the jump in the columnar and orientational order parameters, $C$ and~$R$, is quite
pronounced already for the system sizes studied here, see Figs.\
\ref{op1} and~\ref{op2}. Furthermore, the location the order-parameter
jump provides a good indication of the transition point.
 
 \begin{figure}[t]
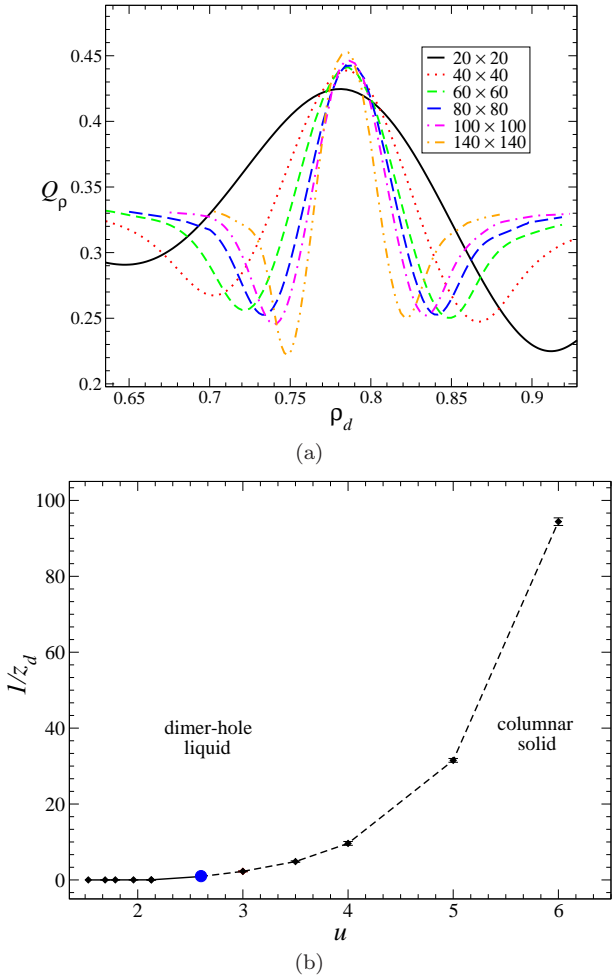

\subfigure[]{\includegraphics[width=0.4\textwidth]{Qdimerd_vs_density_J3.50.eps}\label{dencum}}
\subfigure[]{\includegraphics[width=0.45\textwidth]{fugacity_Jc.eps}\label{zupd}}
\caption{(color online) (a) Fourth-order amplitude ratio for the dimer
density, at coupling $u=3.5$. The exhibited behavior is typical of a
first order transition.\cite{kim03}, with two minima that in the
thermodynamic limit approach the coexisting densities.  (b) Phase
boundary between the dimer liquid and the columnar solid. For low
couplings, the transition is continuous (solid line). Our best estimate
for the multicritical point ($u_c \approx2.6$) is represented by the
large circular dot. Beyond the multicritical point, the phase boundary
(dashed line) is estimated from grand-canonical MC simulations at fixed
\emph{dimer fugacity}~$z_d$.}
\end{figure}

The strongest evidence, however, is provided by the fourth-order
amplitude ratio of the density, $Q_\rho$. At a first-order transition,
this quantity displays a specific behavior, as discussed in
Ref.~\onlinecite{kim03}. In particular, the positions of two minima
observed in Fig.~\ref{dencum} approach, in the thermodynamic limit, the
densities of the two coexisting phases. Outside the coexistence region,
$Q_\rho$ approaches a limiting value~$1/3$, characteristic of Gaussian
fluctuations. This type of behavior is not found at a continuous phase
transition, and should be considered as a strong indicator for the
occurrence of a first-order transition. This is particularly important
since the very weak nature of the first-order transition makes it
impossible to unambiguously confirm the existence of a double peak in
the histograms of the internal energy for the system sizes that we
considered. Whereas the first-order transition becomes more pronounced
at higher couplings, and it thus should become easier to distinguish the
two peaks in the energy histogram, in practice those simulations are
seriously hampered by the very large relaxation times encountered for
large dimer interactions.

By repeating the analysis presented here for different couplings, and
estimating the coexistence chemical potential from the convergence point
observed at the order-parameter discontinuity (cf.\ Fig.~\ref{op}), we
derive the phase diagram in the $z_d$--$u$ plane, see Fig.~\ref{zupd}.

\section{Effects of repulsive hole-hole interactions near the first-order transition region}
\label{sec:hole-int}

We now discuss briefly the effects of additional interactions near the
coexistence curve. It is clear that additional interactions near the
first-order transition line should stabilize more complex ordered
inhomogeneous phases. The simplest interaction that competes with the
tendency of holes to pair and phase separate from the crystal is a weak
nearest-neighbor hole-hole repulsion~$V_h$.  The addition of such an
interaction to the classical dimer model would lead to an additional
energy cost for homogeneous and isotropic clusters of holes.
Remarkably, for a range of dimer interactions~$u$, this energy cost
leads to the formation of commensurate hole stripes with period~$3$, in
a region in the phase diagram located between the dimer-columnar crystal
and the hole-dimer liquid. For general values of $u$, $V_h$ and hole
densities one expects a complex phase diagram, most likely similar to
what is found in theories of commensurate-incommensurate transitions,
which we do not explore here in detail but are discussed in
Refs.~\onlinecite{pokrovsky-talapov,bak82,fisher-selke,papanikolaou07}.

\begin{figure}[h]
\includegraphics[width=0.5\textwidth]{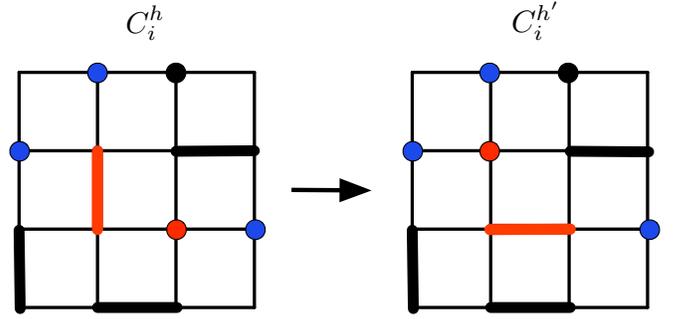}
\caption{(color online) A particular hopping process realized in the
Hamiltonian Eq.~\eqref{Hholeint}. The potential terms that are present
in the Hamiltonian depend on the number of additional pairs of holes
that are formed after the hopping process. In the process shown above,
this number is 1.}
\label{addit_int}
\end{figure}

This phase can be thought of as the ground-state wave function of a
quantum Hamiltonian constructed using the approach described in
Section~\ref{sec:hamiltonians}. The quantum Hamiltonian that leads to
the prescribed wave function includes a generalized form of the
hole-related Hamiltonian of Eq.~\ref{Ham2},
\begin{eqnarray}
  H =H_d + t_{\rm hole}\sum_i \Bigg[- \Big|C^{h}_i\Big>\Big<C^{h'}_i\Big|
  - \Big|C^{h'}_i\Big>\Big<C^{h}_i\Big| \nonumber\\
  + y^{R_{C^{h'}_i}-R_{C^{h}_i}}\Big|C^{h}_i\Big>\Big<C^{h}_i\Big|+y^{R_{C^{h}_i}-R_{C^{h'}_i }}\Big|C^{h'}_i\Big>\Big<C^{h'}_i\Big|\Bigg] \;,
\label{Hholeint}
\end{eqnarray}
where $R_{C_{i}^h}$ and $R_{C_{i}^{h'}}$ denote the number of
pairs of holes formed in the corresponding configurations $C_{i}^{h}$
and $C_{i}^{h'} $ (cf.~Fig.~\ref{addit_int}). More specifically, the ground-state wave function is
\begin{equation}
\ket{
G^{int}_{N_h}}=\frac{1}{\sqrt{Z(w^2,y^2,N_{h})}}\sum_{\{\mathcal{C}_{N_h}\}}
w^{\displaystyle{N^{d}_p[\mathcal{C}_{N_h}]}}y^{\displaystyle{N^{h}_p[\mathcal{C}_{N_h}]}}\ket{\mathcal{C}_{N_h}} \;.
\label{Gzw2int}
\end{equation}
This wave function has a counterpart in the grand-canonical ensemble,
for which the exactly solvable quantum Hamiltonian is a generalization,
in exactly the same way as above, of Eq.~\eqref{Ham3}.

\begin{figure}[hbt]
\includegraphics[width=0.48\textwidth]{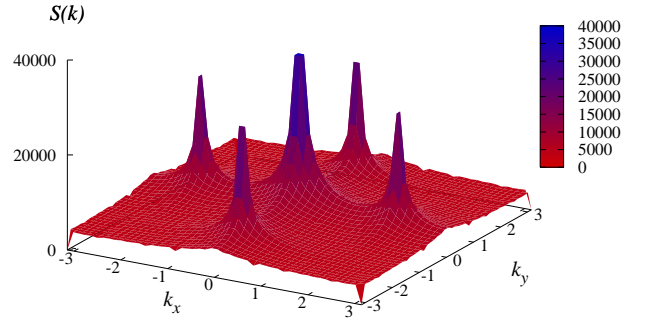}
\caption{(color online) The hole density structure factor~$S(k)$ for
dimer coupling $u=5.0$, hole repulsion $V_h=0.5$ and dimer chemical
potential $\mu_d=-1.0$, for linear system size $L=128$. The four peaks
at $(\pm 2\pi/3,0)$ and $(0,\pm 2\pi/3)$ correspond to the ordered
configuration of hole-stripes shown in Fig.~\ref{cstripe2}.}
\label{holecf}
\end{figure}

\begin{figure}[h]
\includegraphics[width=0.5\textwidth]{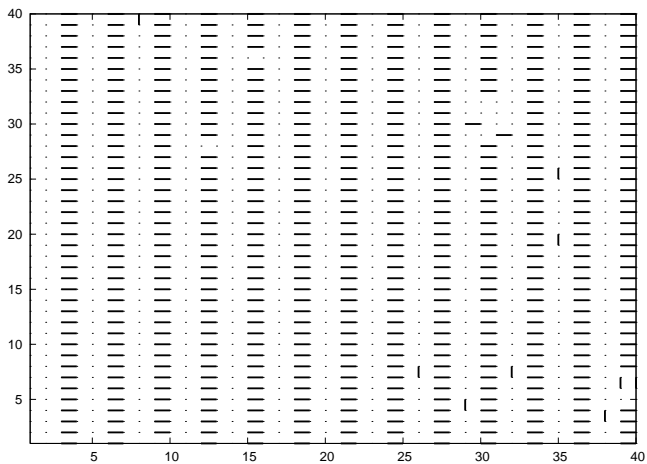}
\caption{Snapshot of a part of a typical ordered configuration that
appears at the couplings $u=5.0$, $V_h=0.5$, $\mu_d=-1.0$ and linear system size
$L=128$. Holes prefer to form commensurate stripes of period $3$, so as
to minimize the effect of the weak hole repulsions.}
\label{cstripe2}
\end{figure}

By performing grand-canonical simulations for weak hole repulsions
$V_h\le\frac{1}{10}u$ in the regime of strong dimer attractions,
$u>4.0$, where the first-order transition is more pronounced, a
hole-stripe phase was observed. In particular, for the couplings
$u=5.0,V_h=0.5$ and a dimer chemical potential $-1.1<\mu_d<-0.8$ (\ie
between the liquid phase $\mu_d \lesssim -1.2$ and the columnar phase
$\mu_d \gtrsim -0.8$), the hole density structure factor shows
non-trivial peaks at $(k_x,k_y)=(\pm 2\pi/3,0),(0,\pm 2\pi/3)$, which
sharpen as the lattice size increases (see Fig.~\ref{holecf}). A
snapshot of the ordered phase (Fig.~\ref{cstripe2}) illustrates how the
holes order in stripes with a period of three lattice spacings, whereas
the dimers are still ordered in a columnar pattern. In this way, the
holes minimize the effect of the hole-hole repulsions and the dimers
simultaneously maximize the effect of the attractive dimer-dimer
interactions. The same pattern is also found for $u=4.0,4.5,6.0$, in
similar regimes for the hole-hole interaction.

More generally, we expect that, as the liquid phase is approached in the
regime of strong dimer couplings, a sequence of hole-commensurate phases
will be stabilized, leading ultimately to incommensurate phases next to
the liquid phase. The formation of this phase diagram is similar in
spirit to the ones discussed in
Refs.~\onlinecite{pokrovsky-talapov,bak82,fisher-selke,papanikolaou07}.

\section{ Elementary quantum excitations of doped quantum Dimer Models}
\label{sec:SMA}

In the preceding two sections we discussed the properties of the
ground-state wave functions and the behavior of equal-time correlation
functions of several operators of physical interest. There we used
extensively the connection that exists for these type of wave functions
between the computation of equal-time correlators of local operators and
computations of similar objects in the equivalent two-dimensional
classical statistical mechanical system of interacting dimers and holes.
In this section we will be interested in the spectrum of low lying
excitations which is inherently a quantum mechanical property.

Unfortunately, as it usually the case in QDMs,\cite{rokhsar88} all we
know is the ground-state wave function. The low-lying excitation
spectrum is not known exactly but it can be computed approximately using
the variational principle. This is the single mode approximation (SMA),
which is a useful tool for studying the excited states of many body
systems.\cite{feynman72,mahan90,arovas91,girvin85,arovas88} It is
particularly useful in the case of quantum dimer models at their RK
points due to the fact that the exact ground-state wave function is
known exactly. The computation of the low lying collective modes in the
QDM was done by Rokhsar and Kivelson.\cite{rokhsar88} Alternatively, one
can describe qualitatively the low-lying spectrum using the effective
field theory of the quantum dimer models (and their generalizations) at
criticality, the {\em quantum Lifshitz model} of
Ref.~\onlinecite{ardonne04}.

In this section we will consider only the SMA spectrum in the dimer-hole
liquid phase. Similar calculations can be done in the phase with
long-range columnar order. In the dimer-hole liquid phase the equal-time
correlation function of the hole density operator, {\it i.e.\/} the
one-body density matrix, approaches a constant at long distances. Thus
the wave function for this phase exhibits a Bose condensate of holes.
Since the holes are charged, this is a charge Bose condensate. To
determine if it is a superfluid (or more precisely a superconductor) it
is necessary to show that it has a finite superfluid stiffness, {\it
i.e.\/} a critical velocity. This can be determined form the spectrum of
density fluctuations and hence from the spectrum of collective modes. It
will turn out that, in spite of the more correlated nature of the wave
functions we consider here, the result will be similar to that of
Rokhsar and Kivelson,\cite{rokhsar88} {\it i.e.\/} no superfluid
stiffness. Given the more general structure (while still local) of the
wave functions we study here, we conclude that wave functions associated
with Hamiltonians satisfying the RK condition in general do not describe
superfluid states.

We begin by summarizing the SMA procedure, focusing on 
doped quantum dimer models. Firstly, one must know the exact ground state
of the system $|0\rangle$ and the type of excitations which 
saturate the frequency sum rule. In our case, there are two candidates:
the dimer density and the hole density excitations. Since it follows
from a variational principle, the SMA provides a proof of existence only
for gapless excitations but not for gapped ones. The energy of an
excitation created by an operator with wave vector $\bf k$, which we
will denote by $\hat\rho_{\bf k}$, acting on the ground state is bounded
from above as follows\cite{feynman72,mahan90,arovas91}
\begin{eqnarray}
E_{\bf k} - E_0 \le \frac{f(\bf k)}{s(\bf
k)}=\frac{\langle0|\left[\hat\rho(-\bf k),\left[\cal{H},\hat\rho(\bf \bf
    k)\right]\right]|0\rangle}{\langle0|\hat\rho(-\bf k)\hat\rho(\bf
k)|0\rangle} \;,
\end{eqnarray}
where $f(\bf k)$ is the ``oscillator strength'' and $s(\bf k)$ is the structure factor ({\it i.e.\/} the equal-time correlation function) for the operator $\hat \rho(\bf k)$.

In the case of doped QDMs at their RK point we know their ground-state
wave functions exactly and they have (by construction) zero ground-state
energy, $E_0=0$. Thus the excitation spectrum must be positive.  The
system will have gapless excitations if $E_{\bf k}-E_0$ vanishes at some
wave vector.  It is worth noting that there are two distinct ways for
the SMA bound to vanish close to some ${\bf k}={\bf k}_0$. One way is if
$f(\bf k)$ vanishes at ${\bf k}_0$. This can happen only if the
commutator $\left[\cal{H},\hat\rho({\bf k}_0)\right]$ vanishes. This
means that $\rho({\bf k}_0)$ is a conserved quantity. The other way
occurs when $s(\bf k)$ becomes infinite at ${\bf k}_0$. This is a
signature of a nearby density-ordered state like the columnar dimer
crystal we found in the phase diagram of the dimer models under study,
shown in Fig.~\ref{fig:sketch}.
 
In the following, we will use the following operator definitions. For
dimers, $\hat\sigma_{\hat \alpha}^{d}({\bf r})$ denotes the dimer
density operator and it takes the values $\pm1$ if a dimer is present or
absent at the link which begins at the position ${\bf r}$ and has
direction $\hat \alpha =\hat x,\hat y$. For holes, $\hat\sigma^{h}({\bf
r})$ denotes the hole density operator and it takes the values $\pm1$ if
a hole is present or absent at the position ${\bf r}$.

Details of the derivations of the SMA oscillator strength functions for
both models are given in Appendix \ref{app:fks}. Here we just quote the
main results.

\subsection{ The fixed hole density model}

\subsubsection{Hole density excitations}

For the fixed hole density model the SMA oscillator strength function
$f(\bf k)$ for hole density excitations is given by
\begin{eqnarray}
f({\bf k})=4t_{\rm hole}{\bf q}^2
\end{eqnarray}
for ${\bf k}=(\pi,\pi)+\bf q$.

From the results of the Section \ref{sec:mean-field} and Appendix
\ref{app:mean-field-details}, we may conclude that: For ${\bf
k}=(\pi,\pi)+{\bf q}$, the structure factor near $(\pi,\pi)$ scales like
$s^{(\pi,\pi)}({\bf k})\propto \frac{1}{q^2+\xi^{-2}}$ and thus,
$s^{(\pi,\pi)}(0)$ is a constant. Thus $E_{\bf k}^{(\pi,\pi)}-E_0\propto
{\bf q}^2$. For ${\bf k}=(0,0)+{\bf q}$, the results from the following
section, which formally hold only for low density of dimers, lead to the
conclusion that $s^{(0,0)}({\bf k})\propto \frac{1}{q^2-\xi^{-2}}$ and
$s^{(0,0)}(0)$ is again a constant but has strong oscillatory behavior
in real space. Thus, from the correlation function we computed in and
Appendix \ref{app:mean-field-details}, we conclude that these
excitations are also quadratic in the momentum ${\bf q}$, \ie\ $E_{\bf
k}-E_{0}\le {\bf q}^2$. This result is consistent with the
compressibility argument, given in Ref.~\onlinecite{rokhsar88} which
would also give quadratic dispersion in this case (keeping in mind that
in this case the compressibility is infinite when the system is doped
(constant number of holes)). However, it is important to stress that the
compressibility argument is a stronger condition because our calculation
of the correlation function is legitimate for low dimer densities.

\subsubsection{Dimer density excitations}

For the fixed hole density model the SMA oscillator strength function
$f(\bf k)$ for dimer density excitations is given by
\begin{eqnarray}
f({\bf k}) = f_{\rm dimer-flip}({\bf k}) + f_{\rm hole(1)}({\bf k})
\label{fkh}
\end{eqnarray}

Close to the wave vector ${\bf Q}_0=(\pi,\pi)$ with ${\bf k}={\bf Q}_0 +
{\bf q}$, both oscillator strengths  $f_{\rm dimer-flip}({\bf k})$ and $f_{hole(1)}$ vanish quadratically (cf. Appendix B) and more specifically,
\begin{eqnarray}
 f_{\rm dimer-flip}({\bf q}) = 8tq^2\\ 
f_{\rm hole(1)}({\bf q})=-t_{\rm hole}q^2
\label{fkh2}
\end{eqnarray}
Given the fact that the dimer density structure factor is a constant at
$Q_0=(\pi,\pi)$ and combining Eqs.\eqref{fkh1} and \eqref{fkh}, we may
conclude that there are gapless dimer density excitations at
$Q_0=(\pi,\pi)$.

In addition to these results, we may have additional branches of dimer
density excitations, specially when there is a divergence of the
structure factor at some wave vector ${\bf Q}_{0}$ which would would be
an indication of dimer order at a nearby phase. In particular, this
happens at the phase boundary between the hole-dimer liquid phase and
the columnar-ordered crystalline phase.

\subsection{The fixed hole fugacity model}

As above, we can study the hole density excitations of the second model,
the grand-canonical model, in which the hole density is not fixed, but
is determined by a parameter $z$ which plays the role of a hole fugacity
in the wave function. By the fact that there is no conservation of the
number of holes is explicitly broken, we can predict that the numerator
$f({\bf k})$ will vanish only at ${\bf Q}_0 =(\pi,\pi)$. This happens
because of the bipartite lattice symmetry that enforces the number holes
on each sublattice to be equal. For the fixed hole fugacity model the
SMA oscillator strength function $f(\bf k)$ for hole density excitations
is
\begin{eqnarray}
f_{\rm hole(2)}({\bf k})=4t_{\rm pairing}(2+\cos({\bf k}_x)+\cos({\bf k}_y))
\label{pairing}
\end{eqnarray}

The formula shows that the numerator $f_{\rm hole(2)}({\bf k})$ vanishes
only at the wave vector ${\bf Q}_0=(\pi,\pi)$ quadratically, as
expected. The dimer density excitations are not gapless (in the sense
that the numerator does not vanish for any ${\bf Q}_0$) because the
dimer-flip contribution vanishes at $(\pi,\pi),(0,\pi)$ and the
dimer-breaking term gives a constant contribution. This is expected due
to the violation of the dimer number conservation.  It is important to
stress that the behavior of the structure factor is exactly the same as
before (in the fixed hole-density model). The reason is the equivalence
of the canonical and grand-canonical ensemble in the thermodynamic limit
for classical systems.

It is certainly worth noting that, although the two ground-state wave
functions of Eqs.\ \eqref{Gzw2} and~\eqref{Gzw3} have the same
ground-state physics, due to the essential equivalence of the classical
canonical and grand-canonical ensembles in the thermodynamic limit, the
nature of their quantum elementary excitations is drastically different.

\section{Conclusions}
\label{sec:conclusions}

In this work we have constructed generalizations of the quantum dimer
model to include the effects of dimer correlations as well as finite
hole doping in the wave function. Throughout we considered the case of
bosonic (charged) holes and neglected the fermionic spinons. We have
constructed generalized RK Hamiltonians whose ground-state wave
functions describe the effects of (attractive) dimer correlations and
finite hole doping.  We have discussed the rich phase diagram and
critical behavior of three doped interacting quantum dimer models at
their RK point using both analytic methods and numerical simulations. We
have shown that the ground-state wave function embodies a complex phase
diagram which consists of dimer-hole liquid and columnar phases
separated by a critical line with varying exponents, ending at a
multicritical point with a Kosterlitz-Thouless structure where the
transition becomes first order. The critical behavior along the low
doping section of the phase boundary was investigated in detail and the
predictions of our scaling analysis were confirmed with large-scale
Monte Carlo simulations.  Monte Carlo simulations were also used to show
that the transition between the dimer-hole liquid and the columnar solid
does indeed become first order, and to estimate the location of the
multicritical point and of the first-order phase boundary.

In the high-doping regime, near the first-order transition line,
additional repulsive interactions among holes were shown to generate, at
the expense of the two-phase region, an even richer phase diagram with
phases in which the dimer-hole system becomes inhomogeneous. In the
regime of strong dimer coupling, $u=5=0.$, and weak hole interactions,
$V_h=0.5$, we found a stripe phase with wave vectors $(2\pi/3,0)$ and
$(0,2\pi/3)$.  In general, we expect the two-phase region to be replaced
by a complex phase diagram with a large number of commensurate and
incommensurate phases. This physics is well known in the context of
two-dimensional classical statistical mechanics of systems with
competing interactions.\cite{pokrovsky-talapov,bak82,fisher-selke}
However, it is interesting to see how it arises at the level of the
exact ground-state wave function of models of strongly correlated
systems, particularly given the current interest on this type of
phenomena in high-temperature superconductors and related
systems.\cite{zaanen89,kivelson93,Kivelson98,emery99,subir-rmp,balents05b}

In this paper we have also presented an analysis of the low-lying
excitations of the quantum system and found that, much as in the case of
the Rokhsar-Kivelson quantum dimer model, the doped system is a
Bose-Einstein condensate but not a superfluid, since the superfluid
stiffness vanishes even though both dimers and holes interact. It is
apparent that in order to render the Bose-Einstein condensate a true
superfluid it is necessary to violate the RK condition which forces the
wave function to have a local structure. The effects of violations to
the RK condition are poorly understood, and we have not investigated
this important problem which is essential to determine the generic phase
diagram of these models.

\begin{acknowledgments}
  We thank C. Castelnovo, C. Chamon, P. Fendley, S.  Kivelson, M.
  Lawler, R. Moessner, C. Mudry, V. Pasquier, P. Pujol, K. S.  Raman, S.
  Sondhi, and M. Troyer for many discussions. This work was supported in
  part by the National Science Foundation through grants NSF DMR 0442537
  (EF), and CAREER Award NSF DMR 0346914 (EL), and by the U.S.
  Department of Energy, Division of Materials Sciences under Award
  DEFG02-91ER45439, through the Frederick Seitz Materials Research
  Laboratory at the University of Illinois at Urbana-Champaign (EL and
  EF). The calculations presented here were in part performed at the
  Materials Computation Center of the Frederick Seitz Materials Research
  Laboratory at the University of Illinois at Urbana-Champaign (EL), the
  Turing XServe Cluster at the University of Illinois and at NCSA
  Teragrid cluster facilities under award PHY060022.
\end{acknowledgments}

\appendix
 \section{Mean-Field Theory for dimers and holes}
\label{app:mean-field-details}

In this Appendix we give details of the mean-field theory summarized in
Section \ref{sec:mean-field}.  We will use Grassmann variable methods to
write down the partition functions for classical interacting and doped
dimers. We will use these methods to derive a simple mean-field theory
for this system. While mean-field theory, as it is well known, fails to
give the correct critical behavior in two-dimensional systems, it offers
a good qualitative description of the phases and, surprisingly, even of
the gross features of the phase diagram. In subsequent sections we will
use more sophisticated analytic and numerical methods to study the phase
transitions.

\subsection{Non-Interacting dimers at finite hole density }
The classical dimer-hole partition function can be formulated in terms
of a Grassmann functional integral, according to the prescription of
Ref.~\onlinecite{Samuel}.
The classical partition function of the dimer problem on any lattice
which is defined by the connectivity matrix $M$ and with fugacity $z$
for the dimers and 1 for the holes is:
\begin{eqnarray}
\mathcal{Z}_{\rm dimer} &=& \int \mathcal{D}\eta\mathcal{D}\eta^\dagger e^{\sum_i \eta_i\eta^{\dagger}_i + \frac{z}{2}\sum_{ij}M_{ij}\eta_i\eta_{i}^\dagger\eta_j\eta_{j}^\dagger}\nonumber\\
&&
\end{eqnarray}

For the square lattice, $M_{ij}$ is 1 if $i,j$ are nearest neighbor
sites and zero otherwise. For the triangular lattice, $M_{ij}$ is 1 if
$i,j$ are nearest neighbors and also next-nearest neighbor sites, but only
along one diagonal, \ie if $j = i + \hat{\bf{x}} +\hat{\bf{y}}$ or $j =
i - \hat{\bf{x}} -\hat{\bf{y}}$.  Also, the fugacity z ranges from 0 to
$\infty$. When $z \rightarrow 0$, the system is filled with holes and
there are very few dimers and when $z\rightarrow\infty$ the system
approaches the close-packed limit with dimers. Both limits are worth of
study due to the existence of an important theorem by Heilmann and
Lieb\cite{lieb1970} which proves under rather general assumptions the
absence of any phase transitions with doping in this model. Thus, the
identification of the phase in the \emph{few dimers} limit is enough to
conclude about the phase that the system enters when doped.  In this
section we will study this problem in the limit in which the dimers are
dilute. In this regime a simple mean-field theory of the Hartree type is
expected to be accurate.\cite{Samuel} Such a crude approximation should
break near criticality, {\it i.e.} near the close packing limit.

To proceed, we apply a Hubbard-Stratonovich transformation to the above
partition function:
\begin{eqnarray}
e^{\frac{z}{2}\sum_{ij}M_{ij}\eta_i\eta_{i}^\dagger\eta_j\eta_{j}^\dagger} &=& \mathcal{N}\int\mathcal{D}\phi e^{-\frac{1}{2z}\sum_{ij}\phi_iM^{-1}_{ij}\phi_j+\sum_i\eta_i\eta_{i}^\dagger\phi_i}
\nonumber \\
&&
\end{eqnarray} 
where $\mathcal{N}$ is an irrelevant normalization constant.  We may
also add sources for the hole density operators $J\eta\eta^\dagger$.  In
this way, the classical dimer partition function may be written as
follows:
\begin{eqnarray}
\mathcal{Z}_{\rm dimer} &=& \int\mathcal{D}\phi\mathcal{D}\eta\mathcal{D}\eta^\dagger e^{-\frac{1}{2z}\sum_{ij}\phi_iM^{-1}_{ij}\phi_j+\sum_i\eta_i\eta_{i}^\dagger(\phi_i + 1 + J_i)}\nonumber\\
&&
\end{eqnarray}
where he have dropped the normalization constant $\mathcal{N}$, as we
will do in what follows. (Note that what makes the above problem
\emph{unsolvable} is the term `1' in the exponent!)

Upon integrating out the Grassmann variables we find:
\begin{eqnarray}
\mathcal{Z}_{\rm dimer} &=& \int \mathcal{D}\phi \; e^{-\frac{1}{2z}\sum_{ij}(\phi_i-J_i)M^{-1}_{ij}(\phi_j-J_j)+\sum_i\ln(\phi_i + 1)}
\nonumber \\
&&
\end{eqnarray}
In the limit $z\rightarrow 0$ we may formulate a legitimate and
well-defined mean-field theory. To this end we rewrite the partition
function in the following way:
\begin{eqnarray}
\mathcal{Z}_{\rm dimer} &=&\int \mathcal{D}\phi \; e^{-\frac{1}{z}\left[\frac12\sum_{ij}(\phi_i-J_i)M^{-1}_{ij}(\phi_j-J_j)-z\sum_i\ln(\phi_i + 1)\right]}\nonumber\\
&=& \int \mathcal{D}\phi \; e^{-\frac{1}{z}\mathcal{S}(\phi)}
\end{eqnarray}
where 
\begin{eqnarray}
S(\phi) &=& \frac{1}{2}\sum_{ij}(\phi_i-J_i)M^{-1}_{ij}(\phi_j-J_j)-z\sum_i\ln(\phi_i + 1)
\nonumber\\
&&
\end{eqnarray}
is the effective action. As $z\rightarrow 0$, we have a theory which has
a well-defined saddle point and the perturbation around this point will
be in powers of $z$ which plays the role of an effective coupling
constant.
In this way, we have a very fast convergent expansion.

The saddle-point is defined as follows:
\begin{eqnarray}
\frac{\delta\mathcal{S}}{\delta\phi_i}\bigg|_{\phi_i=\bar\phi} = 0
\end{eqnarray}
and we take the following equation for $\bar\phi$ :
\begin{eqnarray}
\bar\phi_i = J_i + z\sum_{j}\frac{M_{ij}}{\bar\phi_j + 1}
\end{eqnarray}
In this approximation, the density of holes is given by:
\begin{eqnarray}
\rho_i =\langle\eta_i\eta^{\dagger}_i\rangle &=& \frac{\partial \ln \mathcal{Z}_{dimer}}{\partial J_i}\\
&=& -\frac{1}{z}\frac{\partial \mathcal{S}}{\partial J_i}\\
&=& \frac{1}{\bar\phi_i + 1}
\end{eqnarray}
So, the source $J_i$ in terms of the density of the holes $\rho_i$, is given by:
\begin{eqnarray}
J_i &=& \bar\phi_i - z\sum_{j}\frac{M_{ij}}{\bar\phi_j + 1} \\
      &=& \frac{1}{\rho_i} - z\sum_j M_{ij}\rho_j - 1
\end{eqnarray}
The Legendre transform of the effective action $\mathcal{S}$ is:
\begin{eqnarray}
\Gamma(\rho_i) &=&\frac{1}{z}\mathcal{S}(\bar\phi_j(\rho_i),J_j(\rho_i))+\sum_{i}J_i(\rho_j)\rho_i \\
&=& -\frac{z}{2}\sum_{ij}\rho_iM_{ij}\rho_j + \sum_i\ln(\rho_i)+\sum_i(1-\rho_i) \nonumber \\
&&
\label{LegendreTransform}
\end{eqnarray}
We specialize now to the case of uniform hole density $\rho_i = \rho$
and also for the case of the square lattice where the number of nearest
neighbors is $2D=4$ and assuming that the number of sites on the lattice
is $N$. Then,
\begin{eqnarray}
\Gamma(\rho) &=&  -\frac{z}{2} (4N\rho^2) + N\ln(\rho) + N(1-\rho) 
\end{eqnarray}
At this level of approximation (``Hartree'') the equation of state becomes
\begin{eqnarray}
J = - 4z\rho + \frac{1}{\rho} - 1
\label{eqnstate0}
\end{eqnarray}
So, when the sources are set to zero, the density in terms of the
fugacity, in the limit $z\rightarrow0$, is:
\begin{eqnarray}
\rho\; &=& \frac{2}{1+\sqrt{1+16z}}
\end{eqnarray}
As a check of the approximation, we may expand the result in the region
of $z\rightarrow0$, where the result should be $\rho\simeq1$ for the
hole density:
\begin{eqnarray}
\rho \;
= 1 - 4z +O(z^2)
\end{eqnarray}
 Or, equivalently, for $\rho \to 1^-$,
\begin{eqnarray}
z = \frac{1}{4\rho} \left(\frac{1}{\rho}-1\right) = \frac{1}{4}\left(1-\rho \right)+O((1-\rho)^2)
\label{eqnstate2}
\end{eqnarray}
The hole density-density correlation function can be obtained in the
following way by using the Legendre transform:
\begin{eqnarray}
\mathcal{G}_{ij}&=&\langle\eta_i\eta^{\dagger}_i\eta_j\eta^{\dagger}_j\rangle - \langle\eta_i\eta^{\dagger}_i\rangle\langle\eta_j\eta^{\dagger}_j\rangle\nonumber \\
&=& \frac{\partial^2\ln Z_{dimer}}{\partial J_i\partial J_j} 
= \frac{\partial\rho_i}{\partial J_j}
\end{eqnarray}
and also,
\begin{eqnarray}
\mathcal{G}_{ij}=\left[\frac{\partial^2 \Gamma}{\partial\rho_i\partial \rho_j}\right]^{-1}
\label{invG}
\end{eqnarray}
By using \eqref{LegendreTransform} and \eqref{invG}, we have:
\begin{eqnarray}
  \left[\mathcal{G}_{ij}\right]^{-1}=\frac{\partial^2
  \Gamma}{\partial\rho_i\partial \rho_j} = - zM_{ij}
  -\delta_{ij}\frac{1}{\rho_{i}^2}
\end{eqnarray}
Since $M_{ij}$ is a function of the distance between sites and vanishes
except for nearest neighbors, its Fourier transform is:
\begin{eqnarray}
M(\vec q)&=& a^2\sum_{i}e^{-i{\bf q}\cdot({\bf r}_i - {\bf r}_j)}M({\bf r}_i - {\bf r}_j)\nonumber \\
&=& 2a^2\sum_{\alpha=1}^{2}\cos q_\alpha a
\end{eqnarray}
Also, we set $a=1$ and finally the hole density connected correlation function is:
\begin{eqnarray}
\mathcal{G}(\vec q) &=& \frac{1}{-\frac{1}{\rho^2}-2z\sum_{\alpha=1}^{2}\cos q_\alpha }
\end{eqnarray}
For momenta near ${\bf Q}=(\pi,\pi)$, ${\bf q} ={\bf Q}+{\bf p} $, with ${\bf p}$ small, it becomes
\begin{eqnarray}
\mathcal{G}({\bf p} + {\bf Q})&\simeq& \;-\;\frac{\rho \xi^{-2}}{\xi^{-2} + {\bf p}^2 }
\end{eqnarray}
where $\xi$ is the correlation length
\begin{equation}
\xi =\sqrt{\displaystyle{\frac{1-\rho}{4\rho}}}
\label{eq:corr-length}
\end{equation}
Finally, the connected hole density correlation function (the structure
factor) in real space for $\rho \to 1$ is:
\begin{eqnarray}
\mathcal{G}(r) &= &
\frac{(-1)^{r_x+r_y+1}}{\sqrt{\pi\xi r}} e^{-\xi r}\;\;\;\;\;\;
\label{hhcfn}
\end{eqnarray}
If we restore the units, the correlation length is
$\xi = a\sqrt{\frac{1-\rho a^2}{4\rho a^2}}$.

Surprisingly, given how crude this approximation is, the result of
Eq.~\eqref{hhcfn} is consistent with the numerical results provided by
Krauth and Moessner\cite{krauth03} for much of the dimer density range
they studied. Significant deviations are seen only upon approach of the
close packing regime where the classical dimer model is critical and
this mean-field calculation fails, {\it e.g.\/} the correlation length
diverges as $\rho \to 0$ with exponent $1/2$, given by
Eq.~\eqref{eq:corr-length}, (the mean-field value). The correct value of
the exponent can be deduced from Table \ref{table:0density} and it is
$1/(2-1/4)=4/7$ ($1/4$ being the scaling dimension of the hole operator
for the non-interacting case.) As we show in Section
\ref{sec:zero-density} (and in Table \ref{table:0density}), the
dimension of the hole operator grows from the value $1/4$ for free
dimers to a value of $2$ at the (Kosterlitz-Thouless) transition to the
columnar state, where it should exhibit an essential singularity due to
the marginality of the hole operator.

\subsection{Adding interactions between dimers}

Clearly, the mean-field method for the non-interacting dimer-hole system
can be easily extended to dimer-hole systems with local interactions. In
the case of attractive interactions between parallel dimers, the
partition function of the monomer-dimer system should be
\begin{eqnarray}
\mathcal{Z}_{\rm d-int} = \int \mathcal{D}\eta\mathcal{D}\eta^\dagger \exp\Bigg( \sum_i \eta_i\eta^{\dagger}_i + \frac{z}{2}\sum_{ij}M_{ij}\eta_i\eta_{i}^\dagger\eta_j\eta_{j}^\dagger\nonumber\\ 
+\frac{V}{4}\sum_{ijkl}\tilde M_{ijkl}\eta_i\eta_{i}^\dagger\eta_j\eta_{j}^\dagger\eta_k\eta_{k}^\dagger\eta_l\eta_{l}^\dagger\Bigg)\nonumber\\
\end{eqnarray}
where $V=z^2(e^{u}-1)$ for an attractive interaction of strength $u>0$
between parallel neighboring dimers. $M_{ij}$ represents the
coordination array of the lattice and it takes the value $1$ when $i$ is
nearest neighbor to $j$ and otherwise is zero. In the same respect,
$\tilde M_{ijkl}$ takes the value $1$ only when $i,j,k,l$ are arranged
on a square plaquette and is zero otherwise. The sums run along all
possible lattice sites for each index. We may perform two
Hubbard-Stratonovich transformations by introducing the fields
$\chi,\phi$ and then we have (again, dropping all normalization
constants):
\begin{widetext}
\begin{eqnarray}
\mathcal{Z}_{\rm d-int}&& =\int \mathcal{D}\eta\mathcal{D}\eta^\dagger \mathcal{D}\chi 
\exp\Bigg[\sum_i \eta_i\eta^{\dagger}_i -\frac{1}{V}\sum_{ijkl}\chi_{ij}(\tilde M_{ijkl})^{-1}\chi_{kl}+
 \sum_{ij}\eta_i\eta_{i}^\dagger\eta_j\eta_{j}^\dagger(\chi_{ij}+\frac{z}{2}M_{ij})\Bigg]\nonumber \\
&& =\int \mathcal{D}\eta\mathcal{D}\eta^\dagger \mathcal{D}\chi\mathcal{D}\phi 
 \exp\Bigg[-\frac{1}{V}\sum_{ijkl}\chi_{ij}\tilde M_{ijkl})^{-1}\chi_{kl}
 -\frac{1}{4}\sum_{i}\phi_i(\chi_{ij}+\frac{z}{2}M_{ij})^{-1}\phi_j+ \sum_{ij}\eta_i\eta_{i}^\dagger(\phi_i+1)\Bigg]\;\;\;\;\;
\nonumber \\
&&=\int \mathcal{D}\chi\mathcal{D}\phi  
\exp\Bigg[-\frac{1}{V}\sum_{ijkl}\chi_{ij}(\tilde M_{ijkl})^{-1}\chi_{kl}
-\frac{1}{4}\sum_{ij}\phi_i(\chi_{ij}+\frac{z}{2}M_{ij})^{-1}\phi_j + \sum_{i}\ln(\phi_i+1)\Bigg]
\end{eqnarray}
Just as we did above, we introduce a set of auxiliary sources
$J_{ij}^\chi$ defined on links $ij$, and $J_i^\phi$ defined on sites
$i$. The partition function now reads
\begin{eqnarray}
&& \mathcal{Z}_{\rm d-int}[J^\chi,J^\phi]
=\int \mathcal{D}\chi\mathcal{D}\phi \; e^{\displaystyle{-\frac{1}{Vz}\mathcal{S}(\phi,\chi;J^\phi,J^\chi)}}
\end{eqnarray}
where the action now is:
\begin{eqnarray}
\mathcal{S}(\phi,\chi;J^\phi,J^\chi)&=&z\sum_{ijkl}(\chi_{ij}-J^{\chi}_{ij})(\tilde M_{ijkl})^{-1}(\chi_{kl}-J^{\chi}_{kl})
\nonumber \\
&&+\frac{Vz}{4}\sum_{ij}(\phi_i-J^{\phi}_i)(\chi_{ij}+\frac{z}{2}M_{ij})^{-1}(\phi_j-J^{\phi}_j) 
 -Vz\sum_{i}\ln(\phi_i+1)
 \nonumber \\
 &&
 \label{act-int}
 \end{eqnarray}
 We define the densities conjugate to the fields $\phi_i,\chi_{ij}$ as:
\begin{eqnarray}
n_i = \frac{\partial \ln \mathcal{Z}_{d-int}}{\partial J_i^\phi}=\frac{1}{2}\sum_{j}(\chi_{ij}+zM_{ij})^{-1}\phi_{j}, \qquad
m_{ij}=\frac{\partial \ln \mathcal{Z}_{d-int}}{\partial J_{ij}^\chi}=\frac{2}{V}\sum_{kl}(\tilde M_{ijkl})^{-1}\chi_{kl}
\end{eqnarray}

Solving for the fields and replacing in Eq.\eqref{act-int}, we have finally
for the effective potential or equivalently the Gibbs free energy :
\begin{eqnarray}
\Gamma(z,V)=  \frac{V}{4}\sum_{ijkl}m_{ij}\tilde M_{ijkl}m_{kl}
&+&
\sum_{ij}n_i\bigg(\frac{V}{2}\sum_{kl}\tilde M_{ijkl}m_{kl}+\frac{z}{2}M_{ij}\bigg)n_j
\nonumber \\
&-&
\sum_{i}\ln\Bigg[2\sum_{j}\Bigg(\frac{V}{2}\sum_{kl}\tilde M_{ijkl}m_{kl} + \frac{z}{2}M_{ij}\Bigg)n_j + 1 \Bigg]
\nonumber \\
&&
\end{eqnarray}
The ordered phase of dimers is expected to be a columnar one. So, for
$m_{ij}=(-1)^{x_i}\delta_{i,j-\hat x}m + m_0$ and $n_j=n$, the free
energy is:
\begin{eqnarray}
\frac{\Gamma(z,V)}{N}&=&Vm^2+2Vm_{0}^2+4Vm_{0}n^2+2zn^2
\nonumber \\
&&-\frac{1}{2}\ln\Bigg(1+8n\bigg(V(\frac{m}{2}+m_0)+\frac{z}{2}\bigg)\Bigg)
-\frac{1}{2}\ln\Bigg(1+8n\bigg(V(-\frac{m}{2}+m_0)+\frac{z}{2}\bigg)\Bigg)
\nonumber \\
&&
\label{gibbs}
\end{eqnarray}

Now, we proceed by solving two of the three extremal (saddle-point) equations:
\begin{equation}
\frac{\delta\Gamma}{\delta n}=0 \quad {\rm and} \quad \frac{\delta\Gamma}{\delta m_0}=0
\end{equation}
which take the explicit form
\begin{eqnarray}
  8Vm_0n+4zn- \frac{4\bigg(V(\frac{m}{2}+m_0)+\frac{z}{2}\bigg)}{1+8n\bigg(V(\frac{m}{2}+m_0)+\frac{z}{2}\bigg)} 
  - \frac{4\bigg(V(-\frac{m}{2}+m_0)+\frac{z}{2}\bigg)}{1+8n\bigg(V(-\frac{m}{2}+m_0)+\frac{z}{2}\bigg)}=0 \label{sp1} \\
4Vm_{0}+4Vn^2-\frac{4nV}{1+8n\bigg(V(-\frac{m}{2}+m_0)+\frac{z}{2}\bigg) }
-\frac{4nV}{1+8n\bigg(V(\frac{m}{2}+m_0)+\frac{z}{2}\bigg)}=0 \label{sp2}
\end{eqnarray} 
\end{widetext}
For $z=V=0$, the solution of the saddle-point equations is trivial
$m_0=n=1$. For small $z$, we can solve Eqns.\eqref{sp1} and \eqref{sp2}
recursively, expanding also in the small order parameter $m$ (this is
correct close to a continuous phase transition or to a weakly first
order transition):
\begin{widetext}
\begin{eqnarray}
n&&\simeq1-4z-8V-\frac{1}{4(2V+z)}(\frac{16V^2m^2}{(1+8V+4z)^2}+
\frac{256V^4m^4}{(1+8V+4z)^4}+\frac{4096V^6m^6}{(1+8V+4z)^6})\nonumber \\
&&
\label{n}\\
m_0&&\simeq \frac{2}{1+8V+4z}-1+\frac{1}{4V}(\frac{128V^3m^2}{(1+8V+4z)^3}+
\frac{2048V^5m^4}{(1+8V+4z)^5}+\frac{32768V^7m^6}{(1+8V+4z)^7})\nonumber \\
&& 
\label{m0}
\end{eqnarray}
Replacing Eqs. \eqref{n},\eqref{m0} in the free energy \eqref{gibbs},
and also expanding in the small parameter $z$, we have:
\begin{eqnarray}
\frac{\Gamma(z,V)}{N}=C_0(z,V_r)+C_2(z,V_r)m^2+C_4(z,V_r)m^4
+C_6(z,V_r)m^6 + \cdots
\end{eqnarray}
where $V_r\equiv\frac{V}{z^2}=e^u-1$ and
\begin{eqnarray}
C_0(z,V_r)&&=-2z+(8-2V_r)z^2 
+\frac{32}{3}(-5+3V_r)z^3+O(z^4)\label{c0}\\
C_2(z,V_r)&&=V_rz^2(1+8V_rz^2-128V_rz^3 
-256(-7+V_r)V_{r}z^4)+O(z^7)\label{c2}\\
C_4(z,V_r)&&=-32V_{r}^4z^7(1-2(19+V_r)z
+4(184+V_r(4+V_r))z^2+ 8(1336+
V_r(-232+V_r(4+V_r)))z^{3})  + O(z^{11})\nonumber \\
&&\label{c4}\\
C_6(z,V_r)&&= 128V_{r}^6z^{10}(1-4(V_r+12)z
+4(340+V_r(28+3V_r))z^2)+O(z^{13})\label{c6}
\end{eqnarray}
\end{widetext}

In the limit of very low dimer density, $z\rightarrow 0$, from Eqs.
\eqref{c0}-\eqref{c6}, there is a clear first-order phase transition
from an empty lattice ($n=1,m=0$) to a columnar dimer crystal ($m\neq0$)
because $C_2>0,C_4<0$ and $C_6>0$. In particular, in this limit, the
first-order transition happens when:
\begin{eqnarray}
C_2 &=&\frac{C_{4}^2}{4C_6}\\
z^2V_r &\simeq&\frac{(32V_{r}^4z^7)^2}{512V_{r}^6z^{10}}\\
e^u &=& 1+\frac{1}{2z^2}\simeq \frac{1}{2}z^{-2}\Longrightarrow 2z^2e^u=1 \label{zrw0}
\end{eqnarray}
The condition \eqref{zrw0} can be derived through a very simple
argument: In the limit of very low dimer densities, the non-local
effects which are related to the hard-core dimer constraint are
negligible. If we consider just a single plaquette, then the
contribution from four holes on this plaquette to the Gibbs weight of
the partition function will be just unity but the contribution of two
parallel dimers arranged either vertically or horizontally will be
$2z^2e^u$. When $2z^2e^u=1$, the holes become unstable to the formation
of pairs of parallel nearest neighboring dimers and the result is a
columnar dimer crystal.

When $C_2=C_4=0$ (whereas $C_6$ remains positive), there is a mean-field
tricritical point where the transition ultimately becomes continuous.
Using the approximate Eqs. \eqref{c2}-\eqref{c6}, we can have an
estimate for the tricritical point. Solving the set of equations we
arrive to the following estimate:
\begin{eqnarray}
z_{tr} &=&0.075\\
u_{tr} &=& 2.733
\end{eqnarray}

Remarkably enough, these estimates are very close to the estimates from
the grand-canonical simulations that are presented in Section
\ref{sec:MC}. However, we should be very cautious on taking these
estimates too seriously since the terms in the expansion of the
coefficients $C_0,C_2,C_4,C_6$ in powers of $z$, generically have
alternating signs with increasingly large constant coefficients which
suggests that this expansion is not convergent. In any case, mean-field
approximations, such as the one presented here, fail in two dimensions.
The actual multicritical point has a more complex analytic structure,
akin to a Kosterlitz-Thouless transition, than suggested by these
Landau-Ginzburg type arguments.
On the other hand, the multicritical point can be approached from the
high dimer density limit, where the transition is continuous and thus,
an effective field theory description is possible. As we show in the
following Sections \ref{sec:zero-density},\ref{sec:finite-density}, this
multicritical point is controlled by a marginally relevant operator and
belongs to the Kosterlitz-Thouless universality class.

\section{Derivation of the SMA oscillator strength functions $f(\bf k)$.}
\label{app:fks}

In this Appendix we present the details of the calculations of the SMA
oscillator strength functions $f(\bf k)$ discussed in Section
\ref{sec:SMA}.

\subsection{The fixed hole-density model}
\label{sec:fixed-density}

\subsubsection{Hole density excitations}

The only term of the Hamiltonian \eqref{Ham2} which does not commute
with the hole density operator is the hopping term for the holes. This
term, in terms of destruction-creation Pauli operators
${\sigma_{\hat\alpha}^{d}}^{\pm},{\sigma^{h}} ^{\pm}$ can be written as:
\begin{eqnarray}
  \mathcal{T}_{\rm t-hole}= \hspace{6cm}\nonumber\\
  -\tilde t_{\rm hole}\sum_{<ijk>}{\sigma^{h}}^-({\bf r}_i){\sigma^{h}}^+({\bf r}_k){\sigma^{d}_{\hat\alpha}}^+({\bf r}_{ij}){\sigma^{d}_{\hat\alpha}}^-({\bf r}_{jk})
\end{eqnarray}
At this point, we are not interested for the possible orientations of
the dimers with respect to the holes, as the hole density operator
commutes with the dimer density operator.  Now, we will repeatedly use
that
\begin{eqnarray}
\left[{\sigma^{h}}^{\pm},\sigma^{h}\right] = \mp 2{\sigma^{h}}^{\pm}
\end{eqnarray} 
at the same position in real space. At any distance different from zero,
the commutator vanishes. For a given hole at a position ${\bf R}$, there
are four possible positions ${\bf R}'={\bf R} \pm \hat x \pm\hat y$
where it may move through the available hopping term. By counting
contributions from all these terms for every site of the lattice, we
exactly take into account all the terms of the Hamiltonian including
Hermitean conjugates.

If ${\bf R}$ is the initial hole position and ${\bf R}+{\bf r}_0$ the
final one, then ${\bf r}_0$ has eight possible values: ${\bf r}_0=\pm
\hat x \pm \hat y $.  The first commutator can now be computed for any
site ${\bf R}$(The operator $\mathcal{T}_{\rm hole}({\bf R},{\bf r}_0)$
contains each of the above eight hopping terms). We have:
\begin{widetext}
\begin{eqnarray}
\left[-t_{\rm hole}\mathcal{T}_{\rm hole}({\bf R},{\bf r}_0),{\sigma^{h}}({\bf k})\right] &=&
-t_{\rm hole}\mathcal{T}_{\rm hole}({\bf R},{\bf r}_0)\left(e^{-i{\bf k}\cdot{\bf R}}-e^{-i{\bf k}\cdot({\bf R}+{\bf r}_0)}\right)\\
\left[{\sigma^{h}}(-{\bf k}),-t_{\rm hole}\mathcal{T}_{\rm hole}({\bf R},{\bf r}_0)\right] &=&
 t_{\rm hole}\mathcal{T}_{\rm hole}({\bf R},{\bf r}_0)\left(e^{i{\bf k}\cdot{\bf R}}-e^{i{\bf k}\cdot({\bf R}+{\bf r}_0)}\right)\\
\left[{\sigma^{h}}(-{\bf k}),\left[-t_{\rm hole}\mathcal{T}_{\rm hole}({\bf R},{\bf r}_0),{\sigma^{h}}({\bf k})\right]\right] &=&
2t_{\rm hole}\mathcal{T}_{\rm hole}({\bf R},{\bf r}_0)\left(1-\cos({\bf k}\cdot{\bf r}_0)\right)\;\;\;
\end{eqnarray}
Thus, the oscillator strength $f({\bf k})$ can now be calculated:
\begin{eqnarray}
f({\bf k})&=& \sum_{{\bf R},{\bf r}_0}\langle\left[{\sigma^{h}}(-{\bf k}),\left[-t_{\rm hole}\mathcal{T}_{\rm hole}({\bf R},{\bf r}_0),{\sigma^{h}}({\bf k})\right]\right]\rangle
 2t_{\rm hole}\sum_{{\bf r}_0}\left(1-\cos({\bf k}\cdot{\bf r}_0)\right)
\end{eqnarray}
\end{widetext}
If we set ${\bf k}={\bf Q}_0+{\bf q}$ where ${\bf q}$ is assumed to be small, then: 
For ${\bf Q}_0=(0,0)$, the above expression can be expanded:
\begin{eqnarray}
f({\bf k})&=& 
4t_{\rm hole}{\bf q}^2
\end{eqnarray}
For ${\bf Q}_0=(\pi,\pi)$ we have:
\begin{eqnarray}
f({\bf k})&=& 2t_{\rm hole}\sum_{{\bf r}_0}\left[1-(-1)^{r_{0x}+r_{0y}}\cos({\bf q}\cdot{\bf r}_0)\right]
\end{eqnarray}
$r_{0x}+r_{0y}$ can take only the values $0$ and $2$. Thus, $(-1)^{r_{0x}+r_{0y}}=1$ for all ${\bf r}_{0}$'s. In this way, if we expand around $(\pi,\pi)$, we have again:
\begin{eqnarray}
f({\bf k})=4t_{\rm hole}{\bf q}^2
\end{eqnarray}
It is obvious that for any other ${\bf k}$'s, the oscillator strength $f({\bf k})$ cannot vanish. So, the upper bound to the excitation energy for the holes is:
\begin{eqnarray}
E_{\bf k} - E_0 \le \frac{f({\bf k})}{s({\bf k})}
\end{eqnarray}

\subsubsection{Dimer density excitations}
\label{sec:dimer-density-excitations}

In the same way as before, we may calculate the numerator for the case
where we use the dimer density operator ${\sigma^{d}_{\hat\alpha}}$
instead of the hole density. Now, the operator does not commute with
both the dimer-flip term and the hole-hopping terms.

In the case of the dimer-flip term the commutator will give the
following contribution \cite{rokhsar88, moessner03c}:(the dimer-density
operator is taken to be at $\hat \alpha =\hat x$ direction)
\begin{eqnarray}
f_{\rm dimer-flip}({\bf k})=8t\sum_{{\bf r}=\pm y}\left[1+\cos({\bf k}\cdot{\bf r}_0)\right]
\end{eqnarray}
where $r_0=\pm\hat y$ for the case of a horizontal dimer. This term
comes from the original quantum dimer model. As it was pointed out in
Ref.~\onlinecite{moessner03c}, it vanishes quadratically at
$Q_{0}=(\pi,\pi)$ and $Q_{1}=(0,\pi)$. In particular, at
$Q_0=(\pi,\pi)$, with ${\bf k}={\bf Q}_0 +{\bf q}$, it vanishes as
\begin{eqnarray}
f_{\rm dimer-flip}({\bf q})=8tq^2
\label{fkh1}
\end{eqnarray}

In the case of hole-hopping terms which mix horizontal and vertical
dimers ${\bf r}_0'=-{\bf r}_0/2=(\pm\hat x\pm\hat y)/2$ (where ${\bf
r}_0$ denotes the displacement vector for the hole in the considered
move), we have:
\begin{widetext}
\begin{eqnarray}
\left[-t_{\rm hole}\mathcal{T}_{\rm hole}^{(1)}({\bf R},{\bf r}_0),{\sigma^{d}_{\hat x}}({\bf k})\right]&=&
-t_{\rm hole}\mathcal{T}_{\rm hole}^{(1)}({\bf R},{\bf r}_0)\left(e^{-i{\bf k}\cdot({\bf R}+{\bf r}_0/2)}-e^{-i{\bf k}\cdot{\bf R}}\right)\\
\left[{\sigma^{d}_{\hat x}}(-{\bf k}),\left[-t_{\rm hole}\mathcal{T}_{\rm hole}^{(1)}({\bf R},{\bf r}_0),{\sigma^{d}_{\hat x}}({\bf k})\right]\right] &=&
2t_{\rm hole}\mathcal{T}_{\rm hole}^{(1)}({\bf R},{\bf r}_0)(1-\cos({\bf k}\cdot{\bf r}_0/2))
\end{eqnarray}
\end{widetext}
\begin{eqnarray}
f_{\rm hole(1)}({\bf k})=
2t_{\rm hole}\sum_{{\bf r}_0}(1-\cos({\bf k}\cdot{\bf r}_0/2))
\end{eqnarray}

\subsection{The fixed hole fugacity model}
\label{sec:fixed-fugacity}
Let's follow the same strategy as before (${\bf r}_0=\hat x,\hat y$):
 \begin{widetext}
\begin{eqnarray}
\left[-\tilde t_{\rm hole}\mathcal{T}_{\rm hole}^{(2)}({\bf R},{\bf r}_0),{\sigma^{h}}({\bf k})\right]&=&
\tilde t_{\rm hole}\mathcal{T}_{\rm hole}^{(2)}({\bf R},{\bf r}_0)\left(e^{-i{\bf k}\cdot{\bf R}}+e^{-i{\bf k}\cdot({\bf R}+{\bf r}_0)}\right)
\end{eqnarray}
where $\mathcal{T}_{\rm hole}^{(2)}$ denotes the resonance part of the Hamiltonian in Eq.~\eqref{Ham3}. We also have,
\begin{eqnarray}
\left[{\sigma^{h}}(-{\bf k}),\left[-\tilde t_{\rm hole}\mathcal{T}_{\rm hole}^{(2)}({\bf R},{\bf r}_0),{\sigma^{h}}({\bf k})\right]\right]=
2\tilde t_{\rm hole}\mathcal{T}_{\rm hole}^{(2)}({\bf R},{\bf r}_0)(1+\cos({\bf k}\cdot{\bf r}_0))
\end{eqnarray}
Finally, after adding the contribution of the hermitean conjugate part of the Hamiltonian, we have:
\begin{eqnarray}
f_{\rm hole(2)}({\bf k})=4t_{\rm pairing}(2+\cos({\bf k}_x)+\cos({\bf k}_y))
\end{eqnarray}
\end{widetext}


\end{document}